\documentclass{aa}  

\usepackage{graphicx}
\usepackage{txfonts}
\usepackage{natbib}
\usepackage{pifont}

\usepackage{url}
\bibpunct{(}{)}{;}{a}{}{,} 
\usepackage{color}

\begin{document}

   \title{A solar tornado observed by EIS
}

   \subtitle{Plasma diagnostics}

   \author{P. J. Levens
          \inst{1}
	  \and
	  N. Labrosse
	  \inst{1}
          \and
          L. Fletcher%
	  \inst{1}
	  \and
	  B. Schmieder
	  \inst{2}
          }

   \institute{SUPA School of Physics and Astronomy, University of Glasgow,
              Glasgow, G12 8QQ, UK\\
              \email{p.levens.1@research.gla.ac.uk}
         \and
	Observatoire de Paris, LESIA, 92195 Meudon, France\\
             }

   \date{Received 25 December 2014; accepted 28 July 2015}

 
  \abstract
   {The term `solar tornadoes' has been used to describe apparently rotating magnetic structures above the solar limb, as seen in high resolution images and movies from the Atmospheric Imaging Assembly (AIA) aboard the Solar Dynamics Observatory (SDO). These {often form part of the larger magnetic structure of a prominence, however the links between them remain unclear}. Here we present plasma diagnostics on a tornado-like structure {and its surroundings}, seen above the limb by the Extreme-ultraviolet Imaging Spectrometer (EIS) aboard the Hinode satellite. 
   }
   {We aim to extend our view of the velocity patterns {seen in tornado-like structures with EIS} to a wider range of temperatures and to use density diagnostics, non-thermal line widths, and differential emission measures to provide insight into the physical characteristics of the plasma.}
   {Using Gaussian fitting to fit and de-blend the spectral lines seen by EIS, we calculated line-of-sight velocities and non-thermal line widths. Along with information from the CHIANTI database, we used line intensity ratios to calculate electron densities at each pixel. Using a regularised inversion code we also calculated the differential emission measure (DEM) at different locations in the prominence.}
   {{The {split} Doppler-shift pattern is found to be visible down to {a temperature of} around $\log{\mathrm{T}} = 6.0$}. At temperatures lower than this, the pattern is unclear in this data set. We obtain an electron density {of $\log{n_e} = 8.5$} when looking towards the centre of the tornado structure at a plasma temperature of $\log{\mathrm{T}} = 6.2${, as compared to the surroundings of the tornado structure {where we find $\log{n_e}$ to be nearer $9$}}. Non-thermal line widths show broader profiles at the tornado location when compared to the surrounding corona. {We discuss the differential emission measure in both the tornado and the prominence body, which} {
suggests that there is more {contribution} in the tornado at temperatures below  $\log{\mathrm{T}} = 6.0$ than in the prominence.} }
   {}


   \maketitle
%

\section{Introduction}
\label{sec:intro}

The determination of plasma properties is an essential component of our understanding of the structures observed in the solar atmosphere, and it provides important constraints on the scenarios attempting to explain their properties and appearance {\citep{Labrosse10}}. This work aims to provide an insight into the physical conditions found in the so-called large-scale solar tornadoes that have been observed in the legs of prominences \citep{Wedemeyer13}. There is an ongoing debate about how to interpret the apparent rotation of these features \citep[see][]{2012ApJ...761L..25O,Li12,Su12,panesar13,Su14}. Clearly, combining high-resolution images, spectroscopic observations, and magnetic field measurements is the best approach to making any progress in this discussion. 

Unfortunately, direct measurements of the magnetic field in tornado-like prominences are difficult to perform. It is logical then to combine imaging and spectroscopic data analysis to  understand the nature of these apparent motions. This was done by {\citet{2012ApJ...761L..25O}, \citet{Wedemeyer13}, \citet{Su14}}, amongst others, and their results suggest that the {observed prominence foot} structure  is  rotating. However, their analysis leaves open a few questions related to the state of the plasma that was observed. In particular, the temperature and density of the plasma are two important pieces of information that are needed to put forward and test physical models of tornado-like prominence structures. Similarly, any signature of non-thermal processes in the structure may give a clue as to what types of mechanisms are at work.

In this paper, we investigate the plasma properties of a solar tornado using data obtained by Hinode/EIS and SDO/AIA on 14 September 2013 \citep[also studied by ][]{Su14}. Section~\ref{sec:data} gives an overview of the observations and explains the procedure followed for the data analysis. The analysis of the spectral lines used in this work allowed us to make line-of-sight velocity measurements (Section~\ref{sec:vel}), density diagnostics (Section~\ref{sec:dens}), and a determination of non-thermal line widths in the observed region (Section~\ref{sec:NTLW}). We present our differential emission measure analysis in Section~\ref{sec:DEM}. Section~\ref{sec:conc} gives our conclusions.


\section{Observational overview}
\label{sec:data}

The Extreme-ultraviolet Imaging Spectrometer \citep[EIS][]{Culhane07} aboard the Hinode spacecraft uses a slit (1\arcsec\ or 2\arcsec\ width) or slot (40\arcsec\ or 266\arcsec\ width) followed by a grating spectrometer to pass EUV light to two CCDs (long waveband - 246--292~\AA\ - and short waveband - 170--211~\AA). A few studies have been used to observe prominences with EIS, but very few have focussed on tornadoes. Two related EIS observing plans named \texttt{eis\_tornadoes\_scan} and \texttt{eis\_tornadoes\_sns} were designed by P. G\"{o}m\"{o}ry et al. to look for tornadoes above the limb, and they consisted of a raster and a sit-and-stare observation, respectively. The study made use of the 2\arcsec\ slit of EIS, taking 50 slit positions and covering a spatial extent in $x$ of 100\arcsec. The $y$ extent of the raster was 256\arcsec.

The study was run over {seven} days in September 2013 ({9, 11, 12,} 13, 14, 18 and 19, beginning at around 03:30UT each day) with each day's observing plan consisting of a raster followed by a three hour sit-and-stare, and finally a second raster. One of the seven day's observations (14 Sept.) caught a tornado with the first raster and subsequent sit-and-stare. It is this data set that comprises the focus of this paper. By the second raster of the day, the tornado 
was no longer visible to EIS.

This data set was also used by \citet{Su14}. In that paper the authors discussed the line-of-sight velocities observed in coronal lines formed at around 1.5--2 million K, concluding that the hot plasma observed in both the raster and sit-and-stare indicates that the tornado structure is rotating.

\subsection{Description of the event}
\label{ssec:theevent}
\begin{figure*}
\begin{center}
\includegraphics[scale=0.27,trim=1 0 2 0,clip=true]{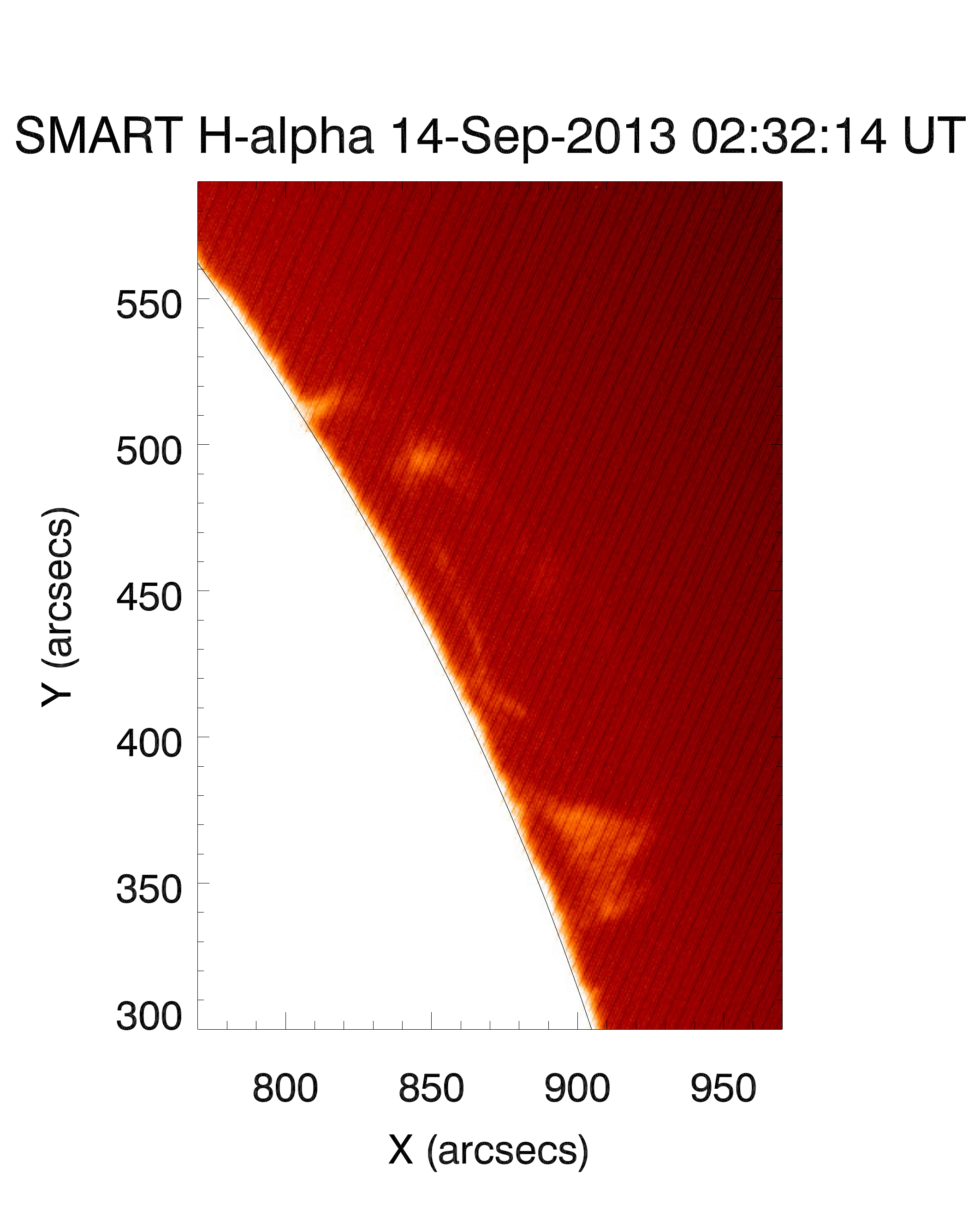}
\includegraphics[scale=0.27,trim=1 0 2 0,clip=true]{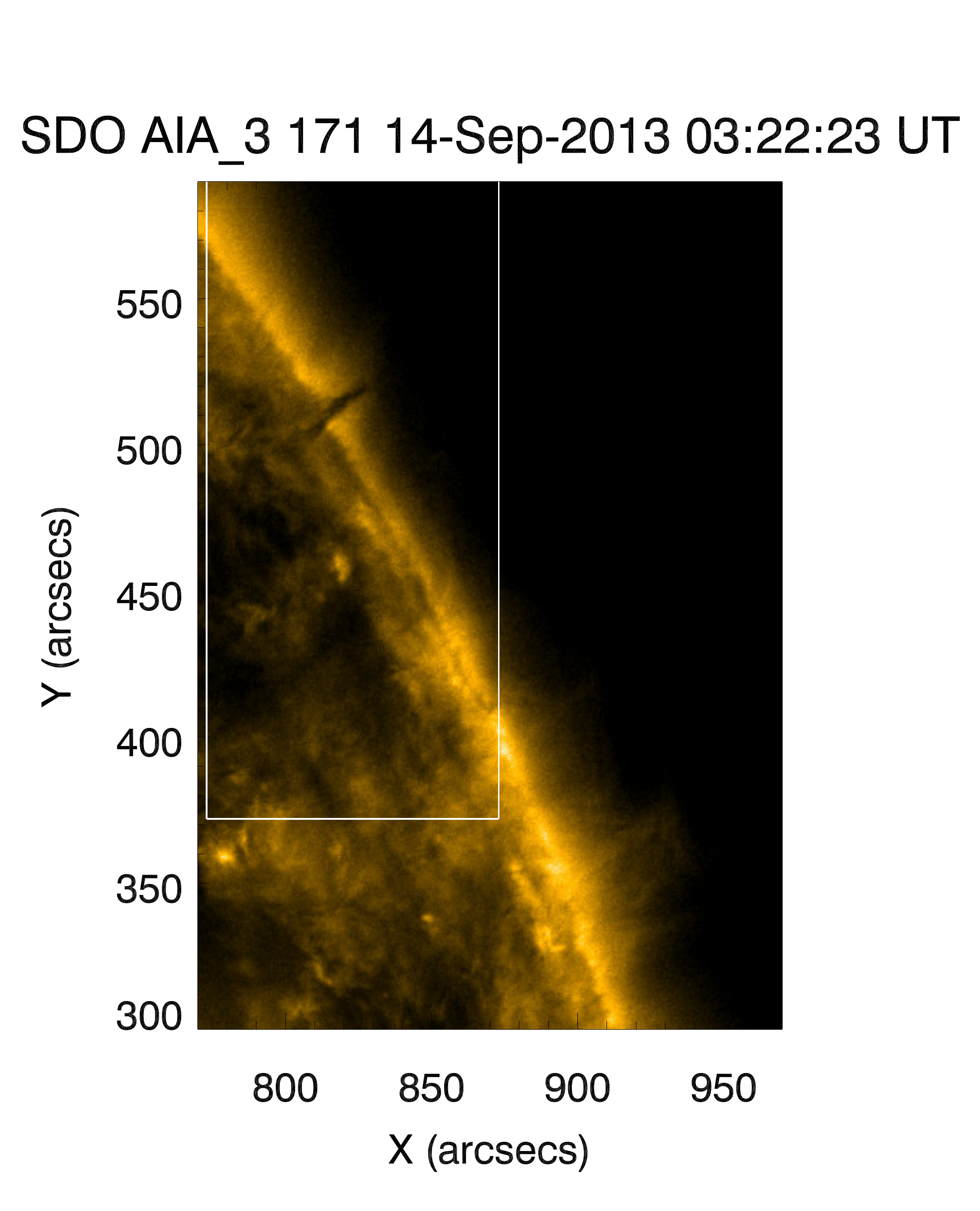}
\includegraphics[scale=0.27,trim=1 0 2 0,clip=true]{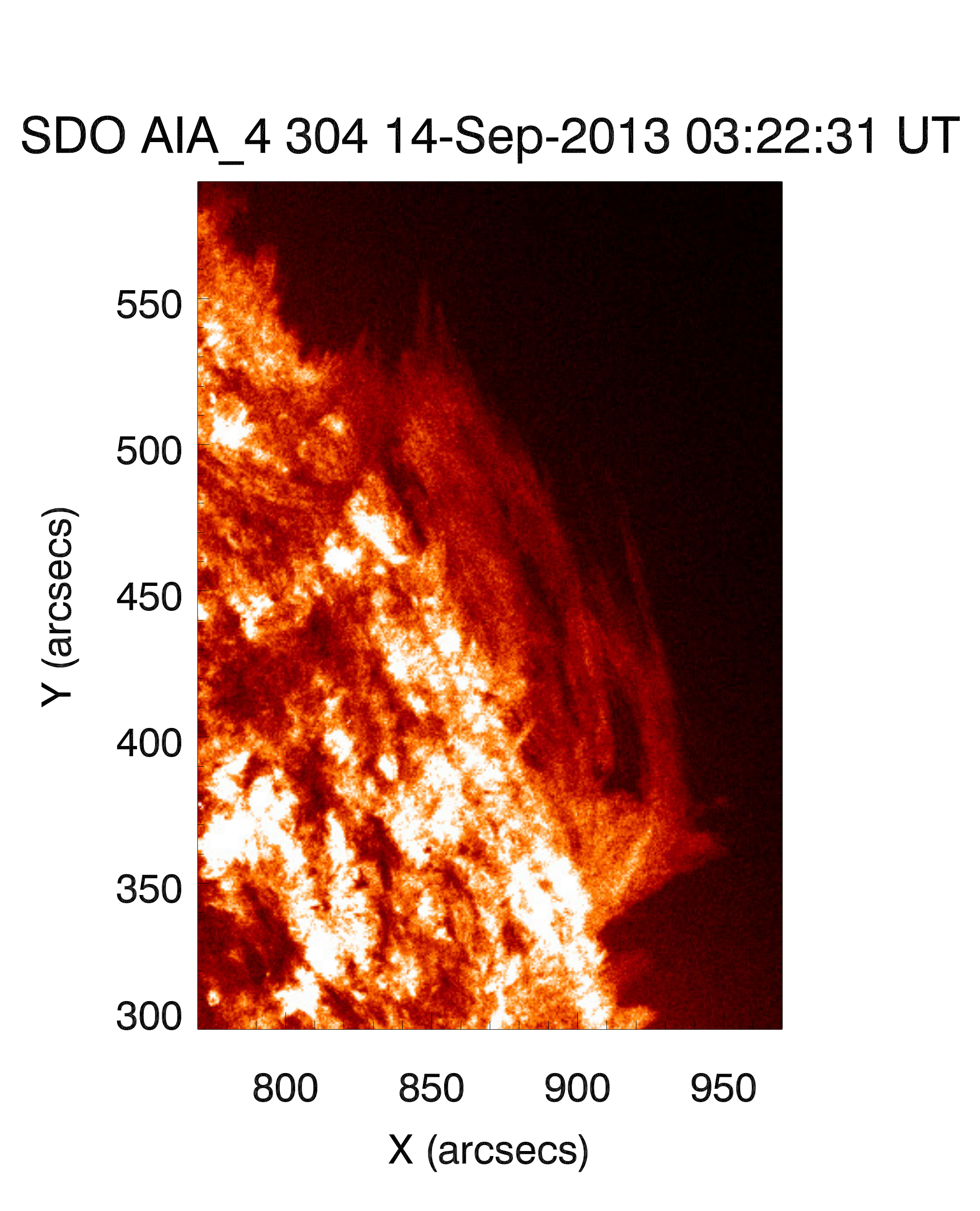}
\caption{Images from the Solar Magnetic Activity Research Telescope (SMART{, Hida Observatory}) and the Atmospheric Imaging Assembly (AIA) aboard the Solar Dynamics Observatory (SDO) showing H-$\alpha$ from SMART (left), the 171~\AA\ passband (centre), and the 304~\AA\ passband (right) from AIA. The central image shows a tornado, absorbing background coronal plasma emission, with the right hand image giving context as to the larger prominence structure. Also shown in white in the central image is the  field of view for the EIS  raster.}
\label{fig:aia}
\end{center}
\end{figure*}

\begin{figure*}
\begin{center}
\includegraphics[scale=0.27,trim=1 2cm 2 2cm,clip=true]{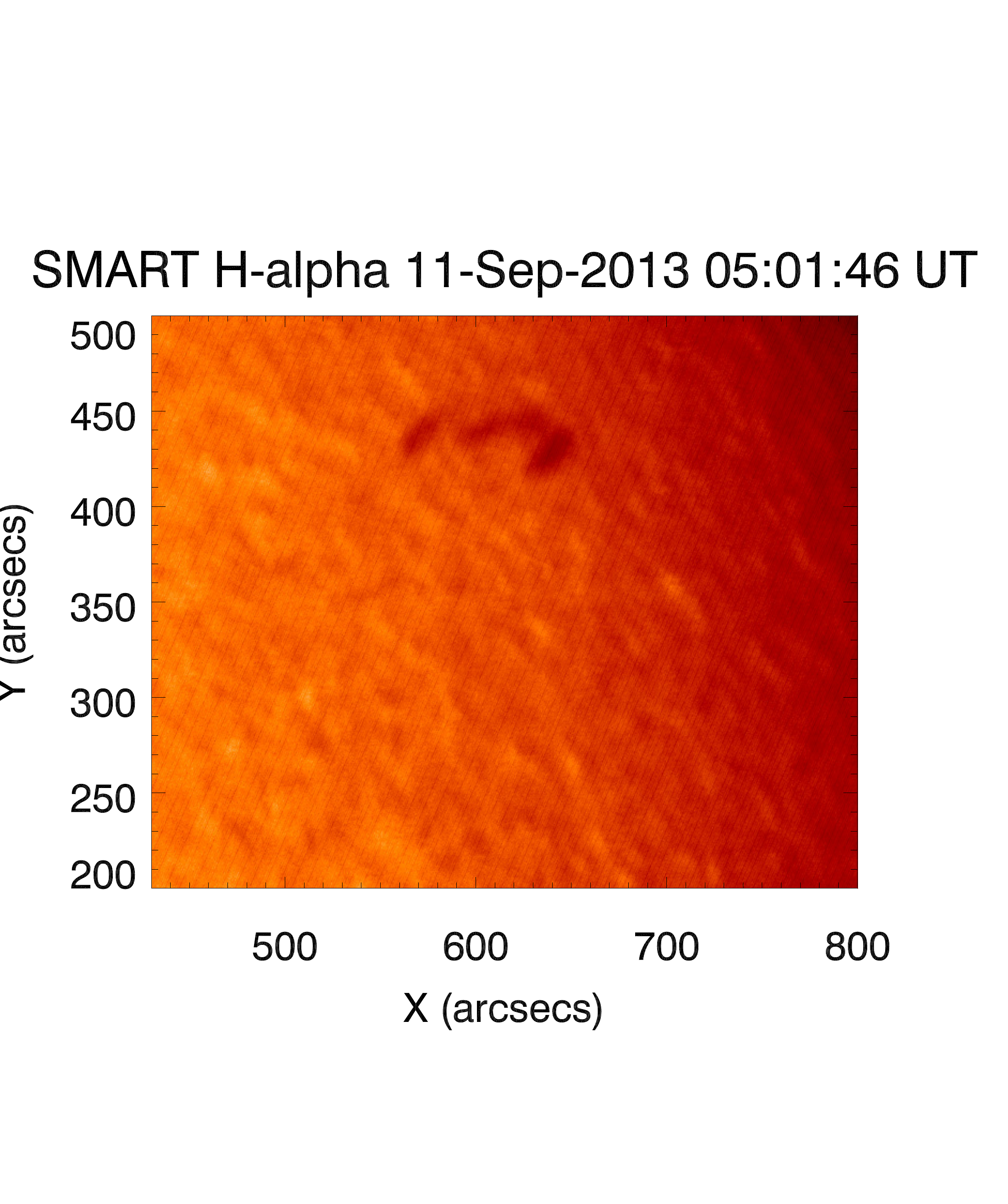}
\includegraphics[scale=0.27,trim=1 2cm 2 2cm,clip=true]{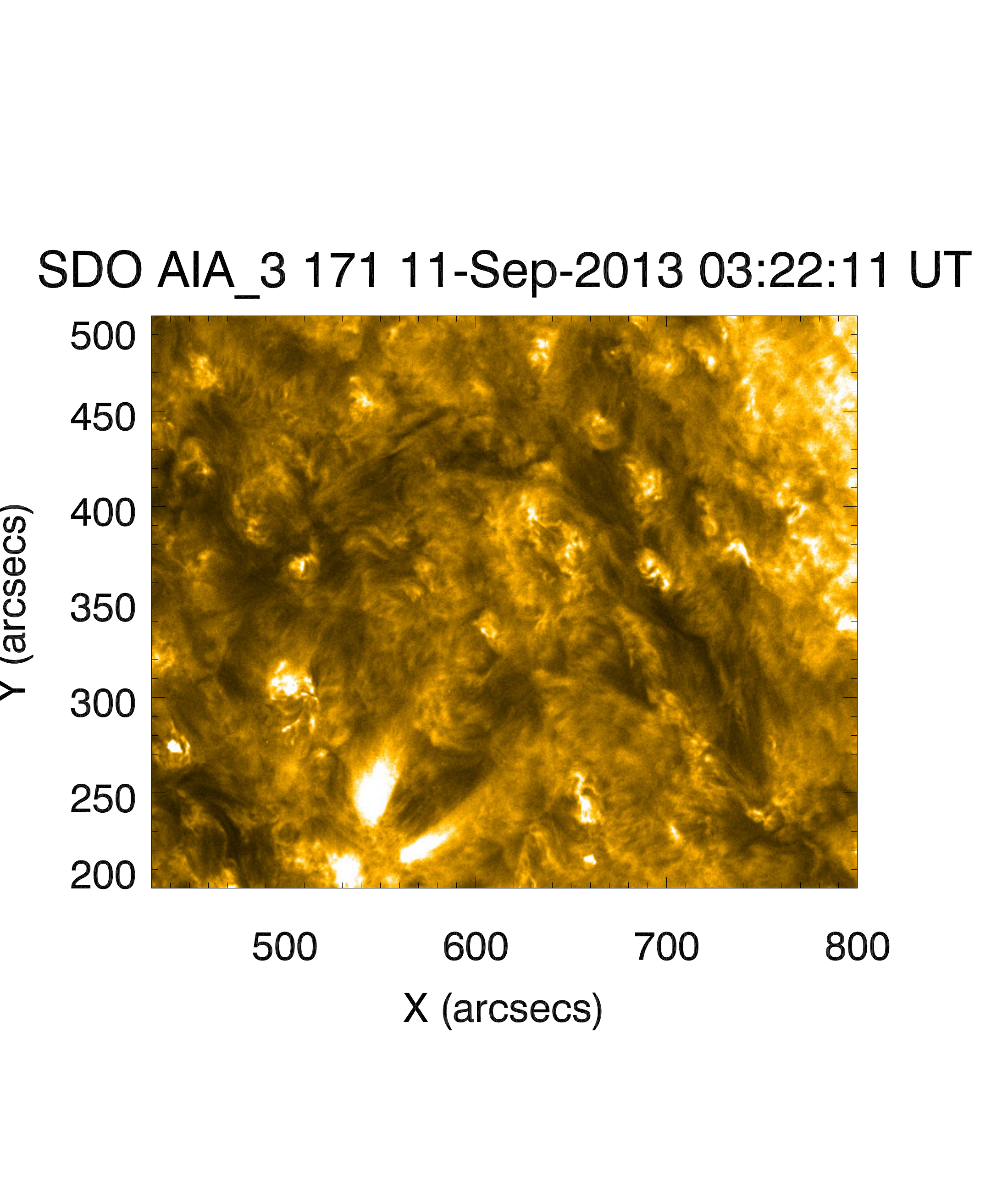}
\includegraphics[scale=0.27,trim=1 2cm 2 2cm,clip=true]{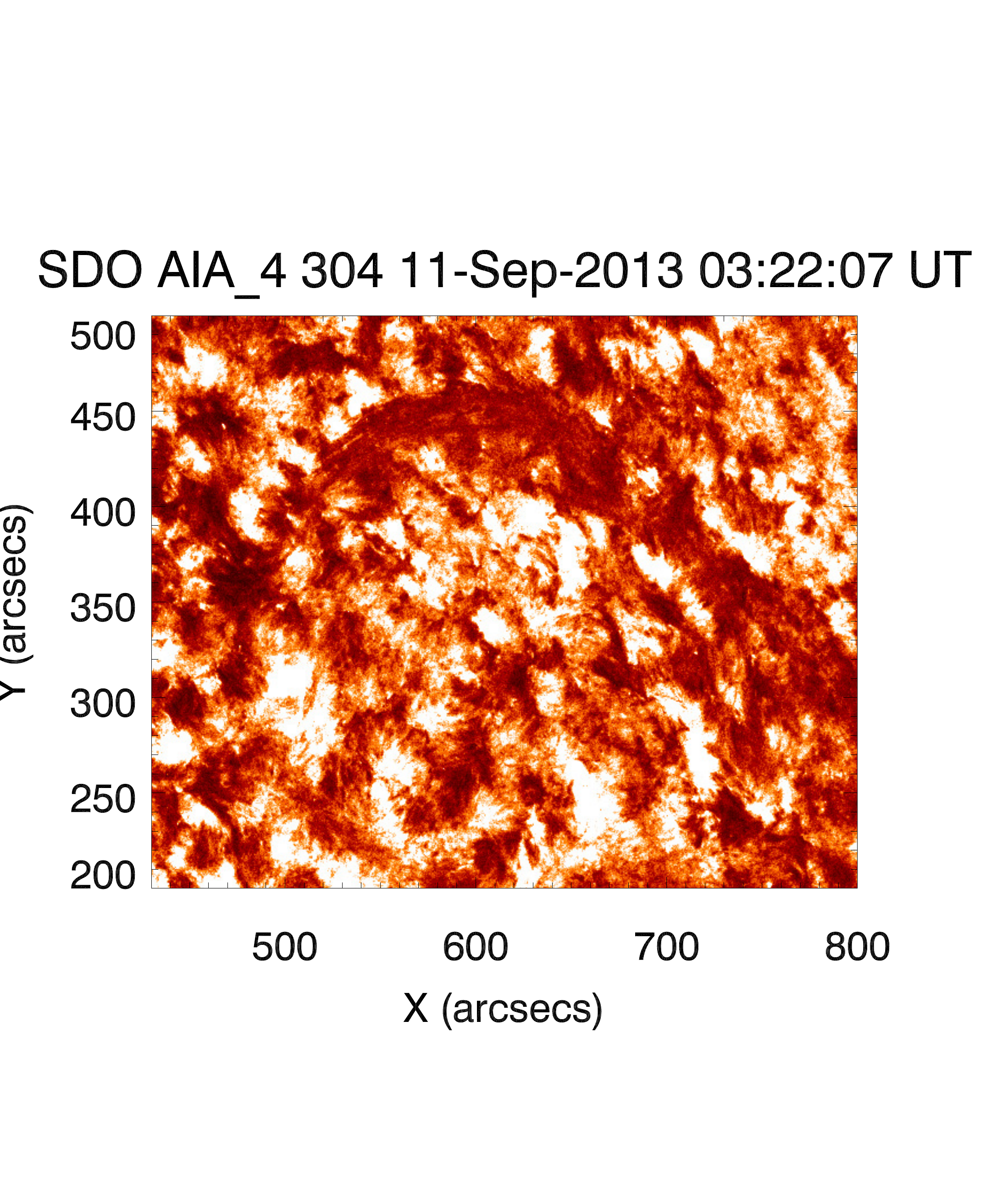}
\caption{Images from SMART H-$\alpha$  (left), AIA 171~\AA\ passband (centre), and AIA 304~\AA\ passband (right). 
}
\label{fig:disc}
\end{center}
\end{figure*}

Figure \ref{fig:aia} shows the tornado as seen in the 171~\AA\ waveband of AIA (central panel), as well as Solar Magnetic Activity Research Telescope (SMART, Hida Observatory, Kyoto University, Japan) H-$\alpha$ (left panel) and the larger prominence structure in AIA 304~\AA\ (right panel).
In coronal lines the tornado is visible to a height of around 20\arcsec\  as a dark feature above the limb. 
~{It appears darker than the surrounding corona because there is a certain volume occupied by cool plasma that is not emitting in hot coronal lines (\emph{emissivity blocking}), and because the cool plasma {absorbs} the radiation coming from behind \citep[see][and references therein]{2008ApJ...686.1383H}.   We know that the dark column on the limb contains hydrogen and helium. The presence of \ion{H}{i}, \ion{He}{i}, and \ion{He}{ii} will lead to absorption of coronal radiation at wavelengths below 912~\AA, 504~\AA, and 228~\AA\ respectively \citep{1976SoPh...50..365O,Labrosse10}. In these dark structures, nearly all background emission in coronal lines in the EIS spectrum is therefore absorbed by the cool plasma along the line of sight \citep{Labrosse11}. Therefore, the emission observed by EIS in the tornado comes from a sheath of hot plasma in front of the tornado and from the corona in front of it.} 
~{Furthermore, }the H-$\alpha$ tornado is similar in size to the dark feature seen in AIA 171~\AA\ images, which is explained by the fact that the optical thickness of the plasma at 171~\AA\ is  comparable to that of  H-$\alpha$  \citep{Anzer05}.
~The prominence 
 can be seen as a filament on disc for a number of days before it crosses the limb (Figure \ref{fig:disc}). 
%

\begin{figure*}
\begin{center}
\includegraphics[scale=0.4]{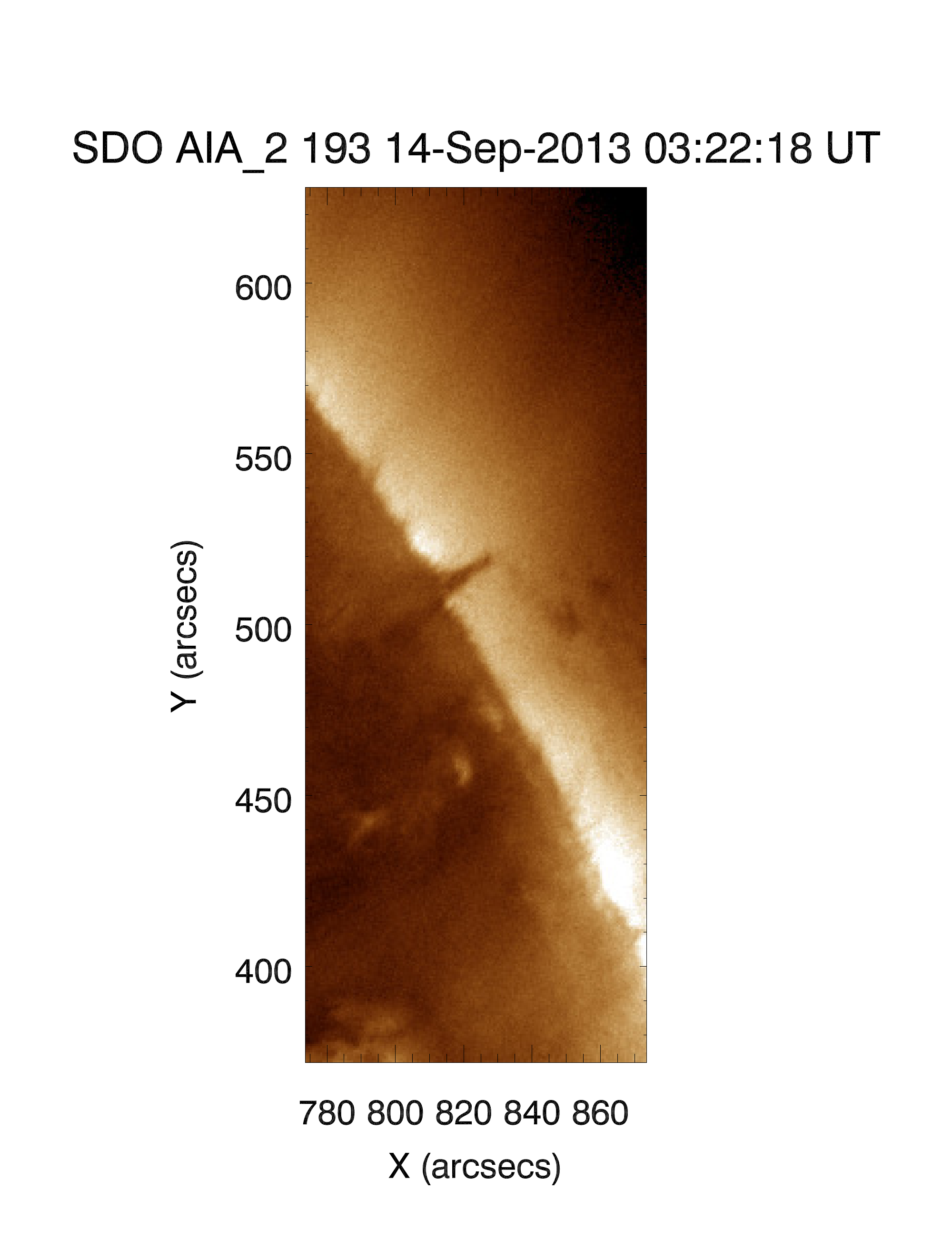}
\includegraphics[scale=0.4]{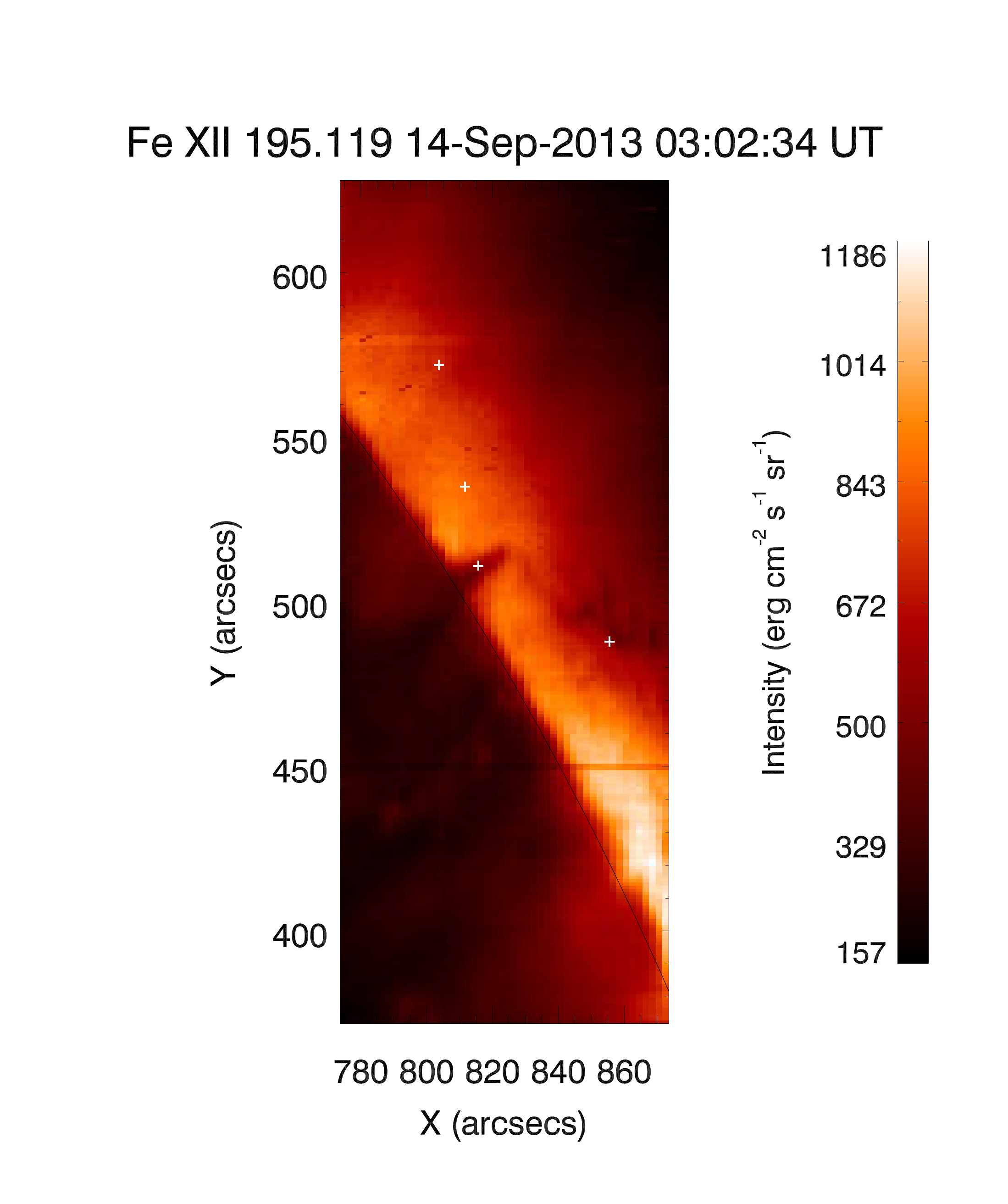}
\caption{{Intensity map{s} of the {AIA 193~\AA\ waveband (left) and the EIS} \ion{Fe}{xii} 195.12~\AA\ spectral line obtained by Gaussian fitting {(right)}, showing the tornado and the structure extending southwards above the limb. The plus markers {in the EIS image (right)} indicate the points at which the {tornado, prominence} {and coronal} DEMs presented in Section \ref{sec:DEM} are calculated. {The AIA image is using the same waveband as the online movie, {that shows the field of view of these images as a white rectangle.}}}}
\label{fig:195int}
\end{center}
\end{figure*}

The EIS raster has a smaller field of view than the AIA and SMART images (Figure \ref{fig:aia}). Figure \ref{fig:195int} shows the intensity image {resulting from the Gaussian fitting} of the 195~\AA\ \ion{Fe}{xii} line {along with an AIA image of the 193~\AA\ waveband. The emission from this AIA window comes predominantly from \ion{Fe}{xii}}. 

{The 195.12~\AA\ line} is the line {in this EIS study} that most clearly shows the tornado structure above the limb.
{In addition to the tornado visible above the limb, this image reveals the presence of a structure extending southwards from the top of the tornado similar to smoke from a chimney, which is co-spatial with the main prominence body seen in 304~\AA\ and with the faint emission seen in H-$\alpha$.}

\subsection{Data reduction}
\label{ssec:datared}
EIS data is aquired as level-0 data files, so it is necessary to run the standard \texttt{eis\_prep} procedure first to bring the data up to level-1. This procedure removes any hot or warm pixels, dark currents, and cosmic ray hits and performs an absolute calibration of the data. {We used CHIANTI \citep{Dere97,Landi12} to identify all the lines visible in the calibrated spectral windows, and fit them assuming Gaussian profiles with the help of Craig Markwardt's \texttt{mpfit} package. The wavelength calibration was carried out using standard routines to correct for orbital variations. We then corrected for solar rotation, choosing {the}
~central {part of} the tornado {to be} fixed {with a} line-of-sight velocity {of} zero.}

It has been noted that the detectors aboard the EIS instrument are {decaying as a function of both time and wavelength \citep{delZanna13, Warren14} instead of simply as a function of time}, so alternative, corrective intensity calibrations must be employed. Two similar methods are well described in those papers, and it is the method outlined by \citet{Warren14} that is used here.

\subsection{EIS spectral lines}
\label{ssec:blend}

{All lines used in this analysis are listed in Table \ref{tab:lines}.} {The emission of these lines is from relatively hot plasma (where $\log{\mathrm{T}} > 5.4$) when compared to the cool plasma of the prominence ($\log{\mathrm{T}} \sim 4$).}
\begin{table}
\centering
\caption{EIS spectral lines used in the analysis}
\label{tab:lines}
\begin{tabular}{c c c c}
\hline\hline
Ion & Wavelength (\AA) & $\log{\mathrm{T(K)}}$ & Blend \\
\hline
\ion{Fe}{x} & 184.537 & 6.1 & \ion{Fe}{xi} \\
\ion{Fe}{viii} & 185.213 & 5.7 & -- \\
\ion{Fe}{xi} & 188.216 & 6.2 & {\ion{Fe}{xii},} \ion{Fe}{xi} \\
\ion{Fe}{xi} & 188.299 & 6.2 & {\ion{Fe}{xii},} \ion{Fe}{xi} \\
\ion{Fe}{xi} & 192.627 & 6.2 & -- \\
\ion{Fe}{xi} & 192.814 & 6.2 & \ion{O}{v}, \ion{Ca}{xvii} \\
\ion{O}{v} & 192.904 & 5.4 & \ion{O}{v}, \ion{Fe}{xi}, \ion{Ca}{xvii} \\
\ion{Fe}{xii} & 195.119 & 6.2 & \ion{Fe}{xii} \\
\ion{Fe}{xii} & 195.179 & 6.2 & \ion{Fe}{xii} \\
\ion{Fe}{ix} & 197.862 & 6.0 & -- \\
\ion{Fe}{xiii} & 202.044 & 6.3 & -- \\
\ion{Si}{vii} & 275.361 & 5.8 & -- \\
\hline
\end{tabular}
\end{table}
{Here we primarily work on the emission spectra observed when looking along the line of sight towards the tornado. According to \citet{Parenti12} this emission is mostly from the prominence-to-corona transition region (PCTR). The word `tornado' here refers to all the hot plasma in the temperature range $\log{\mathrm{T}} = 5.4 - 6.3$ along this line of sight} 

{Due to the finite spectral resolution of the EIS instrument, we observe a number of blended lines. It is possible to resolve some of these blends. The process is described below.}

\subsubsection{\ion{Fe}{x} 184.537~\AA}
\label{sssec:feXblend}
The \ion{Fe}{x} line at 184.537~\AA\ has a small component of \ion{Fe}{xi} in its blue wing. Although much smaller than the \ion{Fe}{x} line itself, this \ion{Fe}{xi} line is non-negligible. Due to the fact that \ion{Fe}{x} dominates the blend, however, a simple two Gaussian fit successfully removed the \ion{Fe}{xi} contribution from the wing.

\subsubsection{The \ion{Fe}{xi} 188.2~\AA\ blend}
\label{sssec:feXI188blend}
The \ion{Fe}{xi} doublet at 188.2~\AA\ is a pair of lines that are partly resolvable, making de-blending them fairly simple{, however there is a small component of \ion{Fe}{xii} in the blue wing of the profile that is not negligible}. A {triple} Gaussian does a good job {of fitting this group}, but the fit can be improved by tying the centroids {of the \ion{Fe}{xi} lines} together. We cannot tie the intensities together as there is a slight density sensitivity between this pair.

\subsubsection{The {\ion{Fe}{xi} and \ion{O}{v}} 192~\AA\ blend}
\label{sssec:caXVIIblend}
This blend at 192~\AA\ represents one of the most complex blends seen by EIS. A number of authors have worked on de-blending these lines \citep[see, e.g.][]{Ko09,Graham13}, mostly in flaring or active region rasters where the \ion{Ca}{xvii} line (formation temperature $\log{\mathrm{T}} = 6.8$) will dominate. In this raster, however, we would expect a minimal amount of emission at this temperature. This changes our approach to de-blending these lines, as the \ion{Ca}{xvii} line will contribute a negligible amount to the blend. Instead, the \ion{Fe}{xi} line becomes the dominant line, with a fairly large amount of \ion{O}{v} emission in this region too.

In this spectral window we also have the \ion{Fe}{xi} line at 192.627~\AA. This line is resolved, outside of the main blend, and can be fitted with a single Gaussian. Although it is weaker than the \ion{Fe}{xi} line within the blend, it is always visible in this raster, and can be used to constrain the centroid of the blended component. As with the 188.2~\AA\ lines discussed in Section \ref{sssec:feXI188blend}, the intensities of these two \ion{Fe}{xi} lines cannot be tied together as they display a density sensitivity.

The five \ion{O}{v} lines in the blend are {difficult} to deal with. These are the only \ion{O}{v} lines available in this study, so we have no way of tying parameters to other \ion{O}{v} lines outside of the blend. These lines were handled by making two assumptions about them. The first is that the five lines can be modelled by two Gaussians centered about the two strongest of the lines - 192.904~\AA\ and 192.797~\AA. The second is that these two Gaussians are tied in both centroid position and in intensity. This is not entirely accurate, as these two lines normally form a density sensitive pair, {however, at the densities considered here this sensitivity is minimal.} As in \citet{Graham13} we here assume a fixed density of {$3 \times 10^{10} ~\mathrm{cm}^{-3}$}, and therefore a fixed intensity ratio of $I_{192.797} = 0.39I_{192.904}$, for these lines. With the \ion{Fe}{xi} line already removed from the blend this is a good approximation for these lines, {even if the assumed density is higher than might be expected for this line - \citet[][table 4 and references therein]{Labrosse10} suggest that the density of \ion{O}{v} would be around $\log{n_e} \sim 9.5$ for an active region prominence, slightly lower for quiescent prominences - the density sensitivity between these two lines at these densities is negligible,} and the resulting fit is sufficient for the analysis performed here.

\begin{figure}
\begin{center}
\includegraphics[width=\hsize]{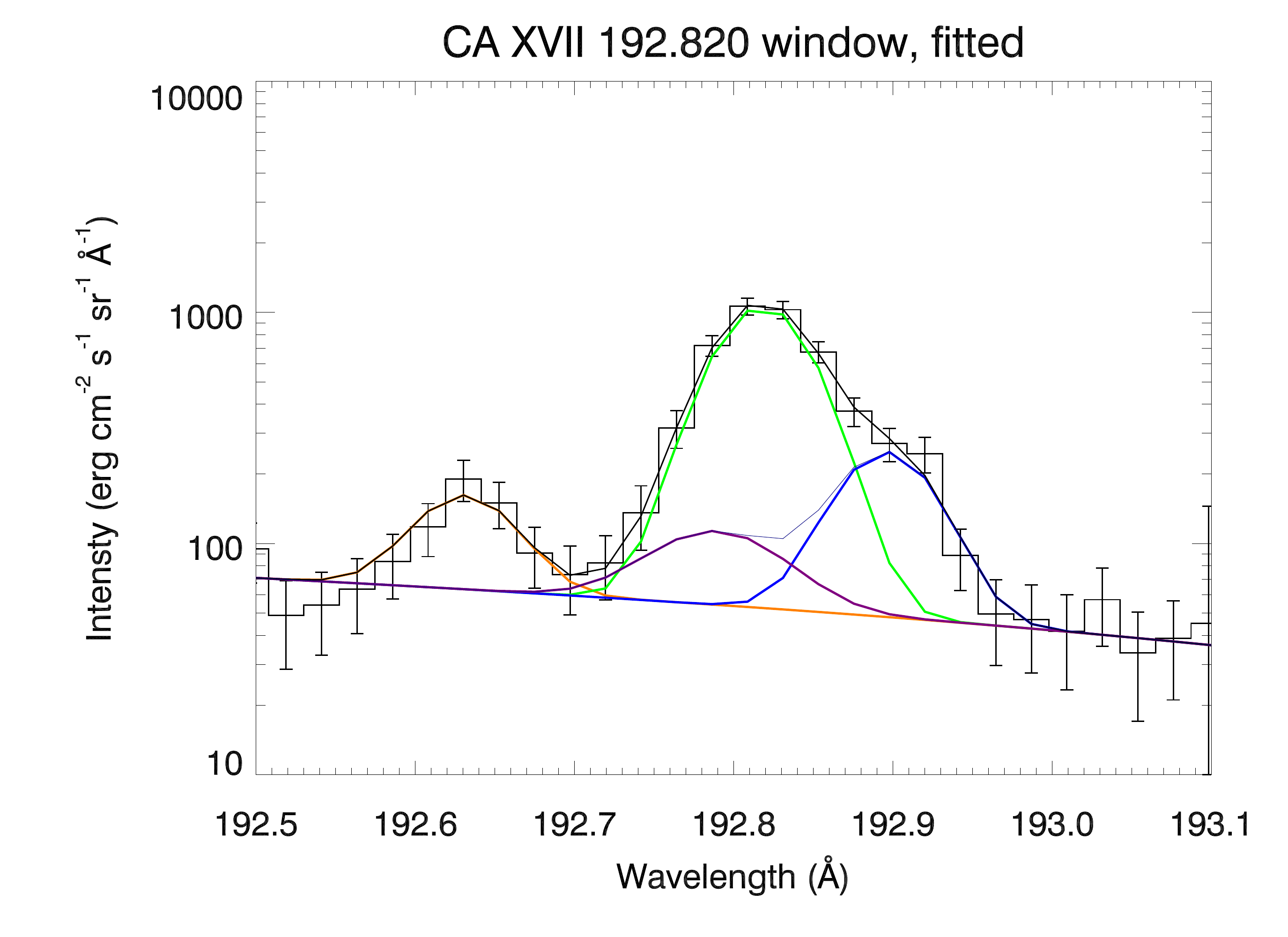}
\caption{{192~\AA\ blend, the most complex blend under consideration in this EIS raster. Shown here is the de-blended line, fitted using four Gaussian components, which are the coloured lines. The original spectrum is shown here as the histogram with error bars. The orange line is the \ion{Fe}{xi} 192.627~\AA\ line, green is \ion{Fe}{xi} 192.814~\AA, and the purple and blue lines represent the \ion{O}{v} lines. The thin, dark blue line shows the overall contribution from the \ion{O}{v} lines, and the solid black line is the total fit to the spectrum. The log scale on the y-axis enhances the weaker lines relative to the stronger lines.}}
\label{fig:192deblend}
\end{center}
\end{figure}

{Figure \ref{fig:192deblend} shows the results of this de-blending process. Seen on the left hand side is the \ion{Fe}{xi}, 192.627~\AA\ line, shown fitted with an orange Gaussian. The fit for the other \ion{Fe}{xi} line is plotted in green, with the two \ion{O}{v} lines in purple and blue. The thin dark blue line represents the total contribution from both \ion{O}{v} lines.}

\subsubsection{The \ion{Fe}{xii} 195.1~\AA\ blend}
\label{sssec:feXIIblend}
\begin{figure}
\begin{center}
\includegraphics[width=\hsize]{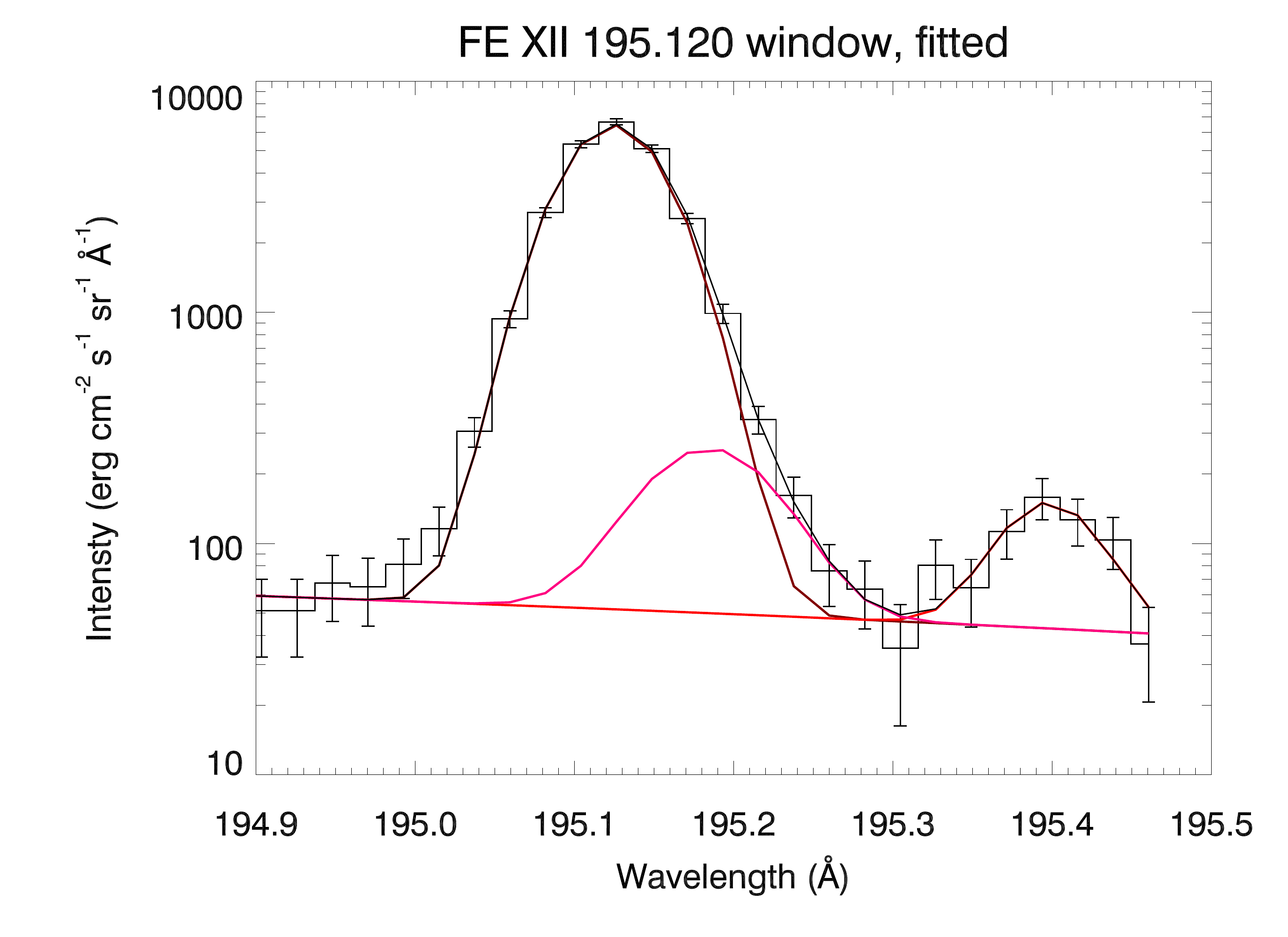}
\caption{{195 \AA\ spectral window as seen by EIS, fitted with multiple Gaussians. Here the histogram with error bars is the raw data as measured by EIS, the dark red curve is the \ion{Fe}{xii} 195.119 \AA\ line and the {magenta} curve is the \ion{Fe}{xii} 195.179 \AA\ component. There is a third Gaussian fitted here, in red, at around 195.4 \AA, that is an unidentified line. The solid black curve is the overall profile created from these Gaussians.} {The log scale on the y-axis emphasises the fits of the weaker lines.}}
\label{fig:195deblend}
\end{center}
\end{figure}

The \ion{Fe}{xii} 195.119~\AA\ line is the strongest line in the EIS spectrum, and is one of the best for analysis of this tornado. In the red wing of this line is a small component of \ion{Fe}{xii} at 195.179~\AA\ that is non-negligible. The 195.119~\AA\ line was one of those used by \citet{Su14} in their analysis of this event, and here we follow a similar approach to remove the red-wing line. We take a double Gaussian fit, and as with the \ion{Fe}{xi} doublet in Section \ref{sssec:feXI188blend} we tie the centroid locations together. Here again we cannot tie the intensities together to form a constant ratio as these two lines form another density sensitive pair. {The fitted 195 \AA\ spectral window is shown in Figure \ref{fig:195deblend}.}

\section{Line-of-sight velocity measurements}
\label{sec:vel}

After all lines visible in the spectral windows were fitted, we used line centroids to calculate the line-of-sight velocity at each pixel. This is the same method used by \citet{Su14}, who {suggest that they found} evidence of rotation of this structure in lines formed at coronal plasma temperatures, specifically in \ion{Fe}{xii} and \ion{}{xiii} lines formed at $\sim$1.5 MK and $\sim$2 MK respectively. It is found, when looking at other lines available in the raster, that this temperature range in which we see this split pattern {of redshifts and blueshifts} along the axis of the structure can be extended down to at least $10^6$~K.

\begin{figure}
\begin{center}
\includegraphics[width=\hsize]{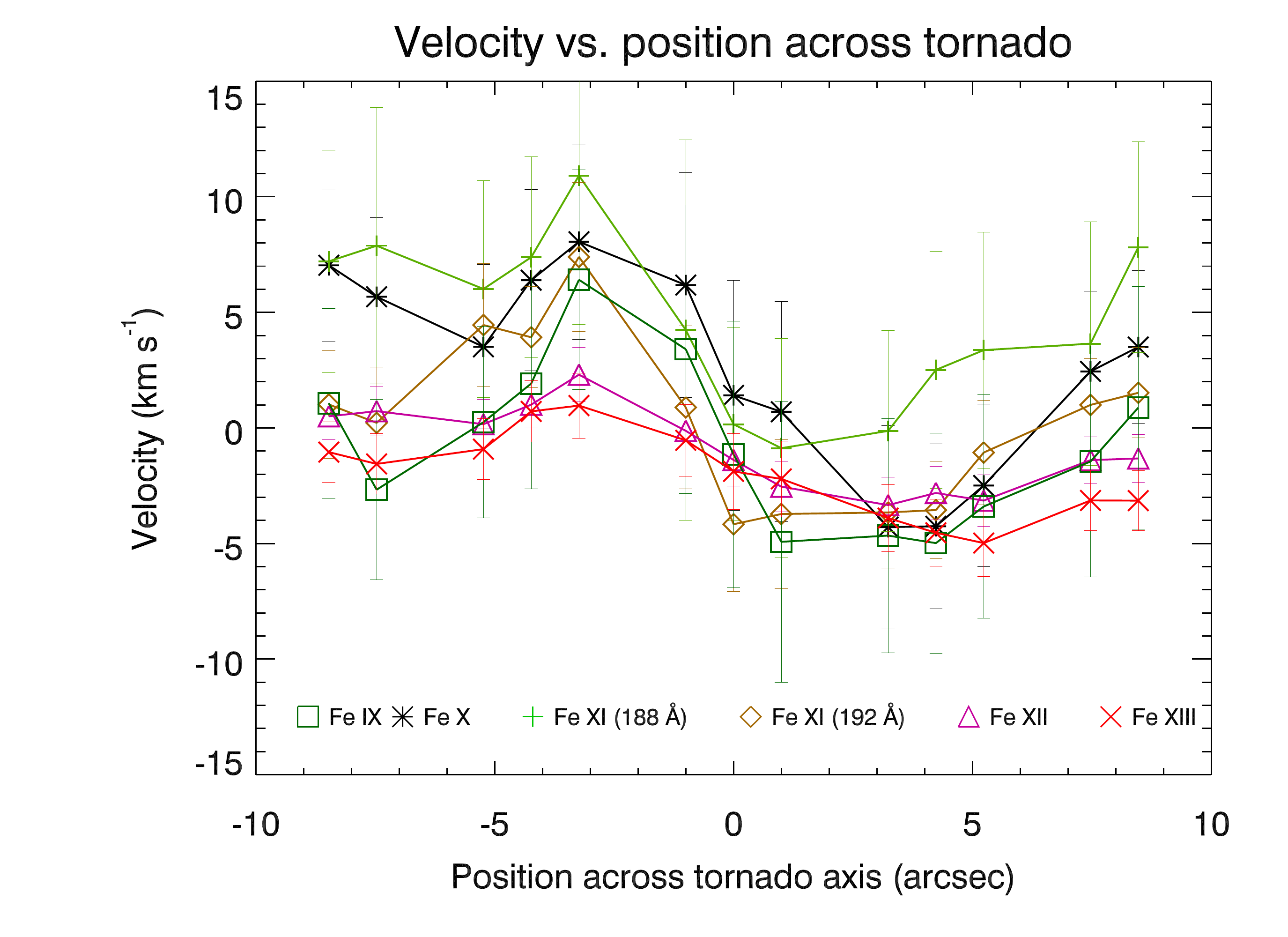}
\caption{{Line-of-sight velocities for \ion{Fe}{ix} -- \ion{Fe}{xiii} ions} for a cut through the tornado -- `cut 3' as seen in Figure \ref{fig:195dens}. The tornado has a spatial extent from approximately [{-5,5}] {arcsec} on the x-axis of this plot.}
\label{fig:fevels}
\end{center}
\end{figure}

Figure \ref{fig:fevels} shows line-of-sight velocities across the tornado for  iron lines {(\ion{Fe}{ix} -- \ion{Fe}{xiii}), with error bars deriving directly from the uncertainties in the fit parameters.} 
The lines used in this plot are all formed at temperatures in the range $\log{\mathrm{T}} = 6.0 - 6.3$. Here we see a Doppler split about the axis of the tornado in each of these lines, with the north side (left half of Figure \ref{fig:fevels}) red shifted and the south (right half of Figure \ref{fig:fevels}) side blue shifted.

\begin{figure*}
\begin{center}
\includegraphics[width=0.45\linewidth,trim=1cm 0 2cm 3cm,clip=true]{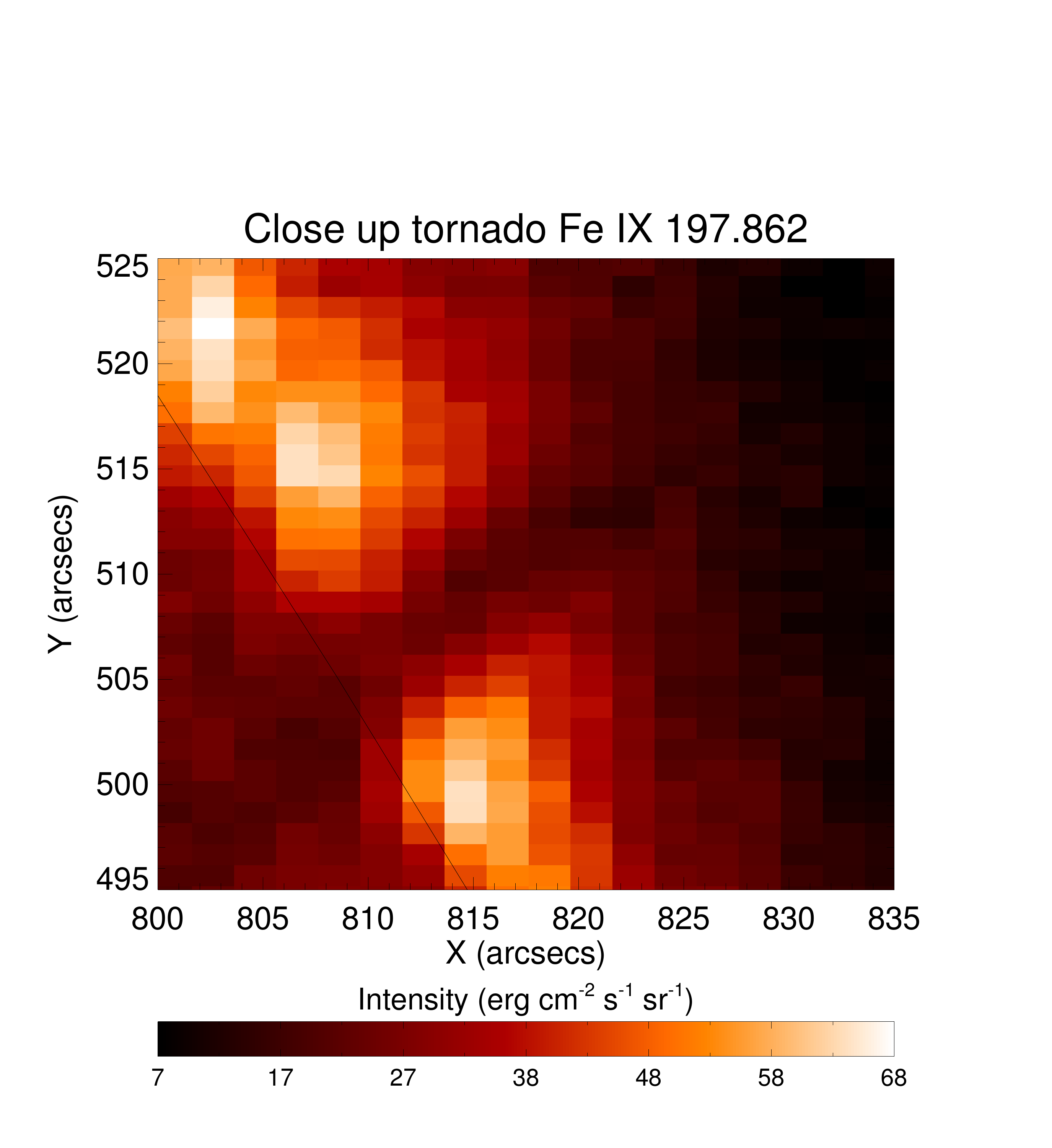}
\includegraphics[width=0.45\linewidth,trim=1cm 0 2cm 3cm,clip=true]{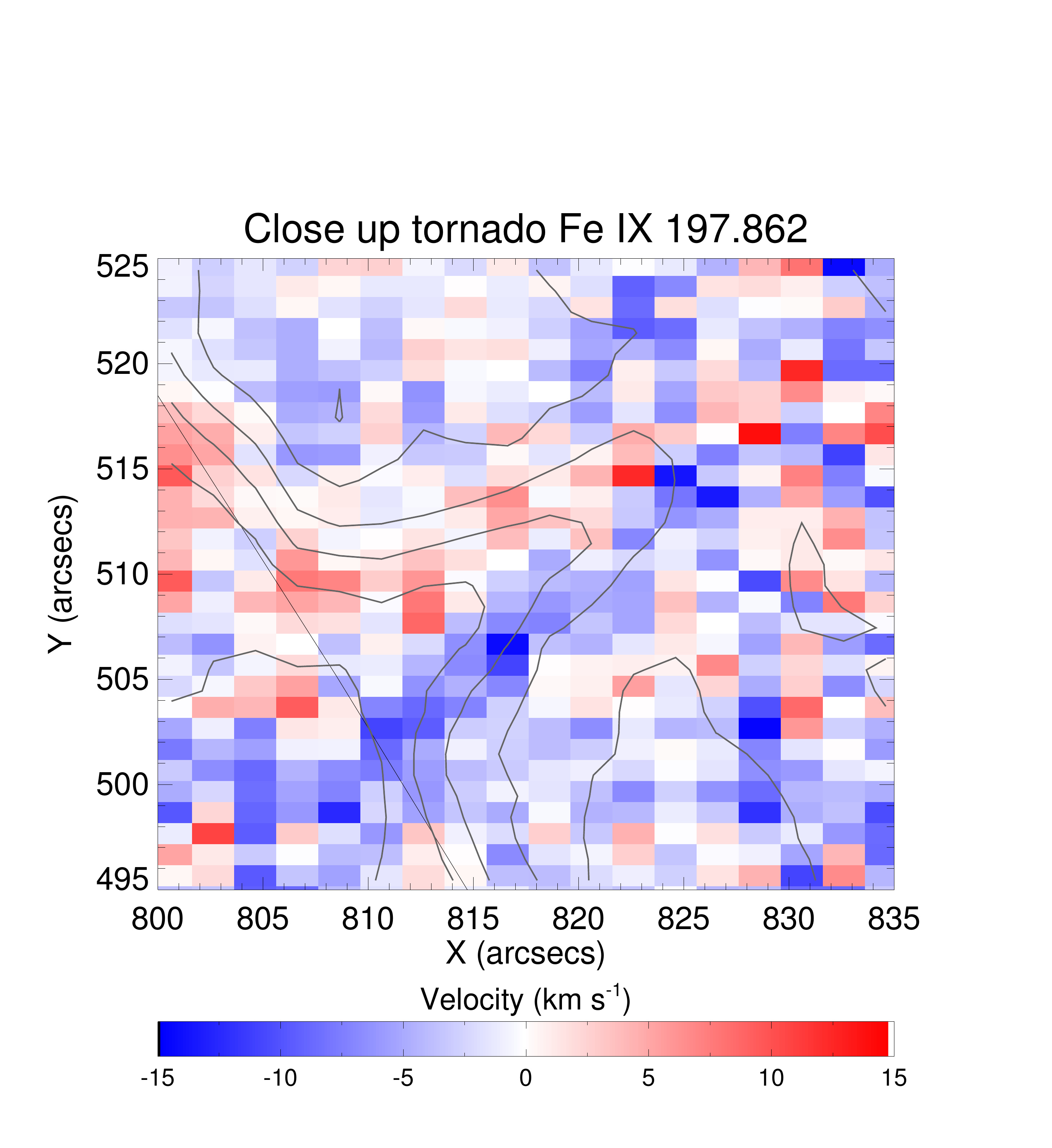}
\caption{Intensity (left panel) and line-of-sight velocity (right panel) maps for {the} \ion{Fe}{ix} 197.86~\AA\ line. {The} contour plot on the velocity map is the \ion{Fe}{xii} 195~\AA\ line {intensity} (Figure \ref{fig:195int}), that shows the tornado structure most clearly above the limb.}
\label{fig:197vel}
\end{center}
\end{figure*}

The \ion{Fe}{ix} line at 197.86~\AA\ is formed at $\sim$$10^6$K, and although the pattern is less clear, we see in Figure \ref{fig:197vel} the same red/blue pattern that is described in the \citeauthor{Su14} paper for hotter plasma temperatures.
%

\begin{figure}
\begin{center}
\includegraphics[width=\hsize,trim=1cm 0 0 3cm,clip=true]{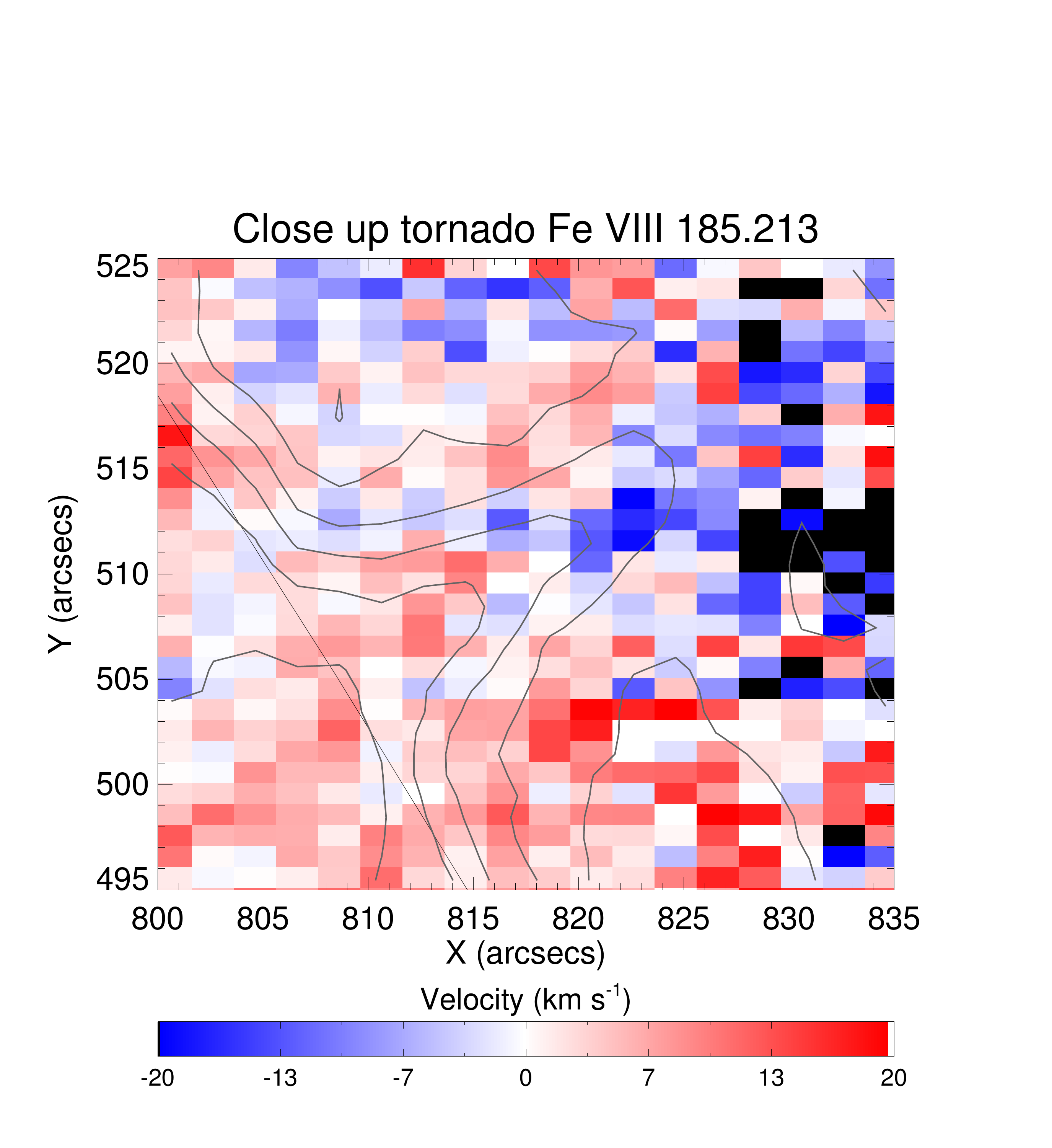}
\caption{Line-of-sight velocity map for the \ion{Fe}{viii} 185.21~\AA\ line. The contour plot is the \ion{Fe}{xii} 195~\AA\ line intensity (Figure \ref{fig:195int}).}
\label{fig:185vel}
\end{center}
\end{figure}

We cannot see clear evidence of this Doppler pattern {for other lines in the study that are formed at temperatures less than $\sim$1 MK}. For example the \ion{Fe}{viii} line at 185.21~\AA, formation temperature of $\log{\mathrm{T}} = 5.7$ and shown in Figure \ref{fig:185vel}, shows no signs of the patterns visible at higher plasma temperatures.
This cannot, however, be taken as evidence that there is no split Doppler pattern at these temperatures, as all lines available in this temperature range {are affected by blends and/or low signal-to-noise ratio (SNR). For example, the \ion{O}{v} lines are heavily blended, (see Section \ref{sssec:caXVIIblend}). In order to see any systematic line shifts we need these to be accurately de-blended, as the typical line-of-sight velocities are small (on the order of a few km~s$^{-1}$). Low SNR for some lines above the limb also make it difficult to measure line-of-sight velocities accurately. This is especially evident in the \ion{Si}{vii} 275.36~\AA\ line (formation temperature of $\log{\mathrm{T}} = 5.8$) and in the \ion{Fe}{viii} 185.21~\AA\ line.}

{One effect that we must acknowledge, that could be responsible for the observations of the split Doppler pattern at higher temperatures, is an artificial Doppler shift that may be introduced by a point spread function (PSF) that is both tilted and elliptical, similar to that seen in \textit{SOHO}/CDS data \citep{Haugan99}.
\ According to \citet[][Appendix B]{Young12} additional Doppler shifts may be found in EIS spectra in regions where there is a large intensity gradient north-south along the slit. This instrumental line shift is typically offset by about three to four pixels from the intensity maximum ({EIS Wiki: {http://solarb.mssl.ucl.ac.uk:8080/eiswiki/Wiki.jsp?page=Spatial \%20offset\%20of\%20intensity\%20and\%20velocity\%20features}}). When there is an increasing intensity gradient from north to south, a blueshift is introduced, whereas decreasing intensity gradients north to south introduce redshifts. In either direction it is found that the additional Doppler velocity is around 5 km~s$^{-1}$.

If this instrumental line shift is present in this data set, then the observations presented here could be seen as spurious.
\ However, we see in the data presented in Figure \ref{fig:197vel}, especially in the top half of the tornado, that there is not a large intensity gradient, yet the Doppler pattern persists. Also, {as can be seen in Figure \ref{fig:fevels},} in the \ion{Fe}{x} {and \ion{Fe}{xi}} line{s} we recover Doppler shifts {exceeding} $\pm$ {5} km~s$^{-1}$, so above the level of the estimated instrumental shift.}


\section{Density diagnostics}
\label{sec:dens}

\begin{figure}
\begin{center}
\includegraphics[width=\columnwidth]{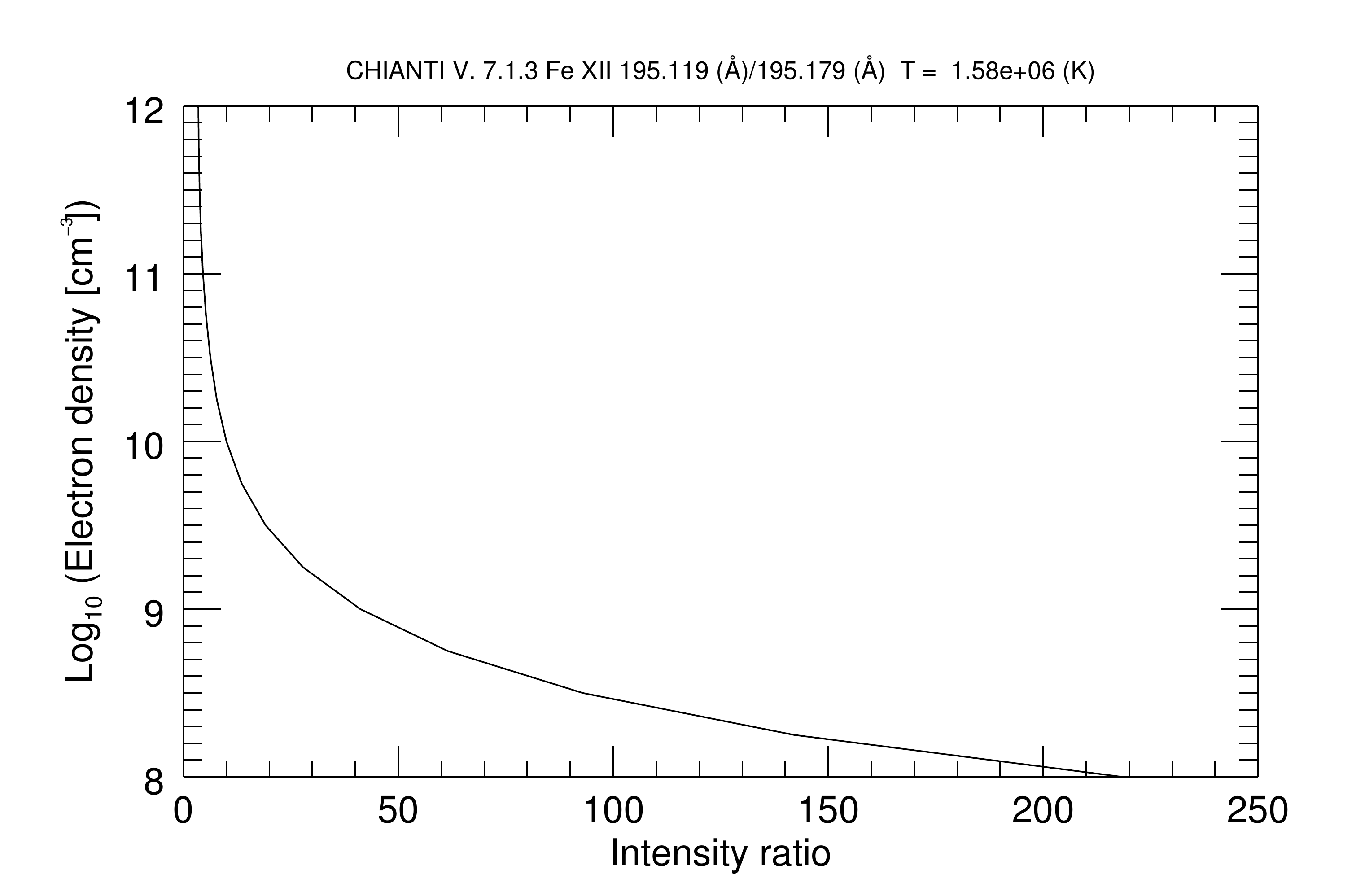}
\caption{Electron density versus intensity ratio for the \ion{Fe}{xii} 195.119~\AA/195.179~\AA\ line pair.}
\label{fig:195densrat}
\end{center}
\end{figure}

There are a number of density sensitive lines visible to EIS, and a few of those were available in this study. Using the CHIANTI {package} \citep{Dere97,Landi12}, that has information on expected intensity ratios for density sensitive line {pairs}, we can calculate the electron density at each pixel position. Figure \ref{fig:195densrat} shows an example density curve for the \ion{Fe}{xii} doublet 195.119~\AA\ and 195.179~\AA.

There are four {available} density diagnostic pairs in this {study}, three from \ion{Fe}{xi} and \ion{Fe}{xii}, all {corresponding to temperatures} around $\log{\mathrm{T}} = 6.2$. A potential \ion{O}{v} diagnostic was ruled out because of the {difficulties in} fitting the {blended} lines, as described in Section \ref{sssec:caXVIIblend}.

\subsection{\ion{Fe}{xi} diagnostics}
\label{ssec:fe11dens}
\begin{figure}
\begin{center}
\includegraphics[width=\hsize,trim=0 0 0 3cm,clip=true]{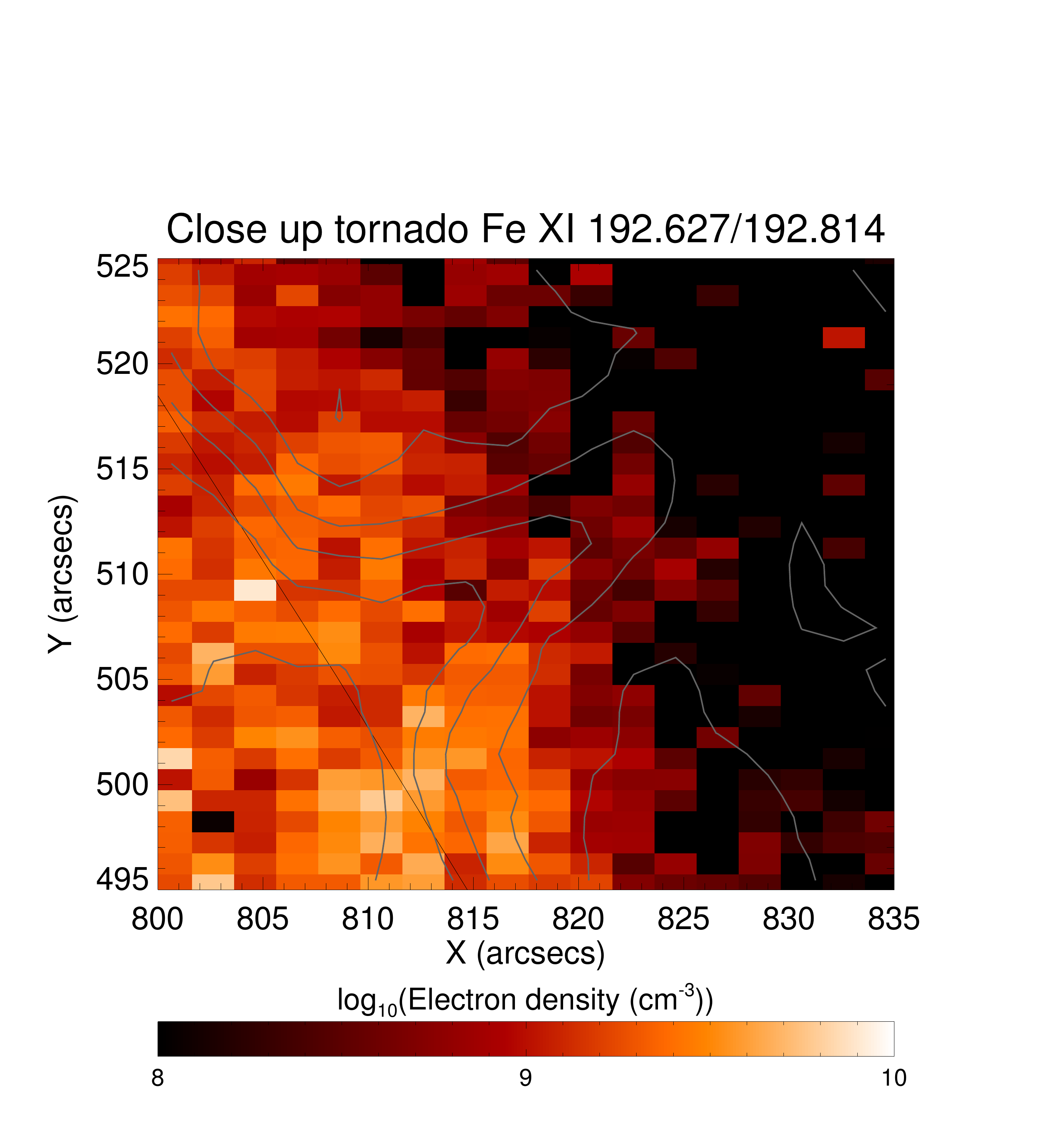}
\caption{Density map for one of the \ion{Fe}{xi} diagnostics. The other two available provide a very similar result. These are not shown here, but are consistent with the diagnostic presented.}
\label{fig:192dens}
\end{center}
\end{figure}

The three available \ion{Fe}{xi} diagnostics all make use of the 192.627~\AA\ line as one of the diagnostic pair. The other three lines used in the diagnostics are each of the 188~\AA\ pair and the 192.814~\AA\ line. There are no useful mutual diagnostics between the three of these, meaning we only have three diagnostic pairs in total for this ion. In each of these, however, we see a slight dip in density at the tornado location. This is seen in Figure \ref{fig:192dens}, which shows one of these \ion{Fe}{xi} diagnostics with the \ion{Fe}{xii} 195.119~\AA\ intensity contour plotted on top. 
The other two density maps provide a similar result, showing a lower density region when looking towards the tornado centre.

\subsection{\ion{Fe}{xii} diagnostic}
\label{ssec:fe12dens}
The \ion{Fe}{xii} density diagnostic available consists of the doublet at 195~\AA. {This is a blend that is dominated by the 195.119~\AA\ line, with the weaker line centred at 195.179~\AA\ contributing a non-negligible amount to the red wing of the blend.} Using the fitting techniques outlined in Section \ref{sssec:feXIIblend}, these lines can be de-blended and used for the diagnostic.

\begin{figure*}
\begin{center}
\includegraphics[scale=0.4,trim=0 3cm 0 3cm,clip=true]{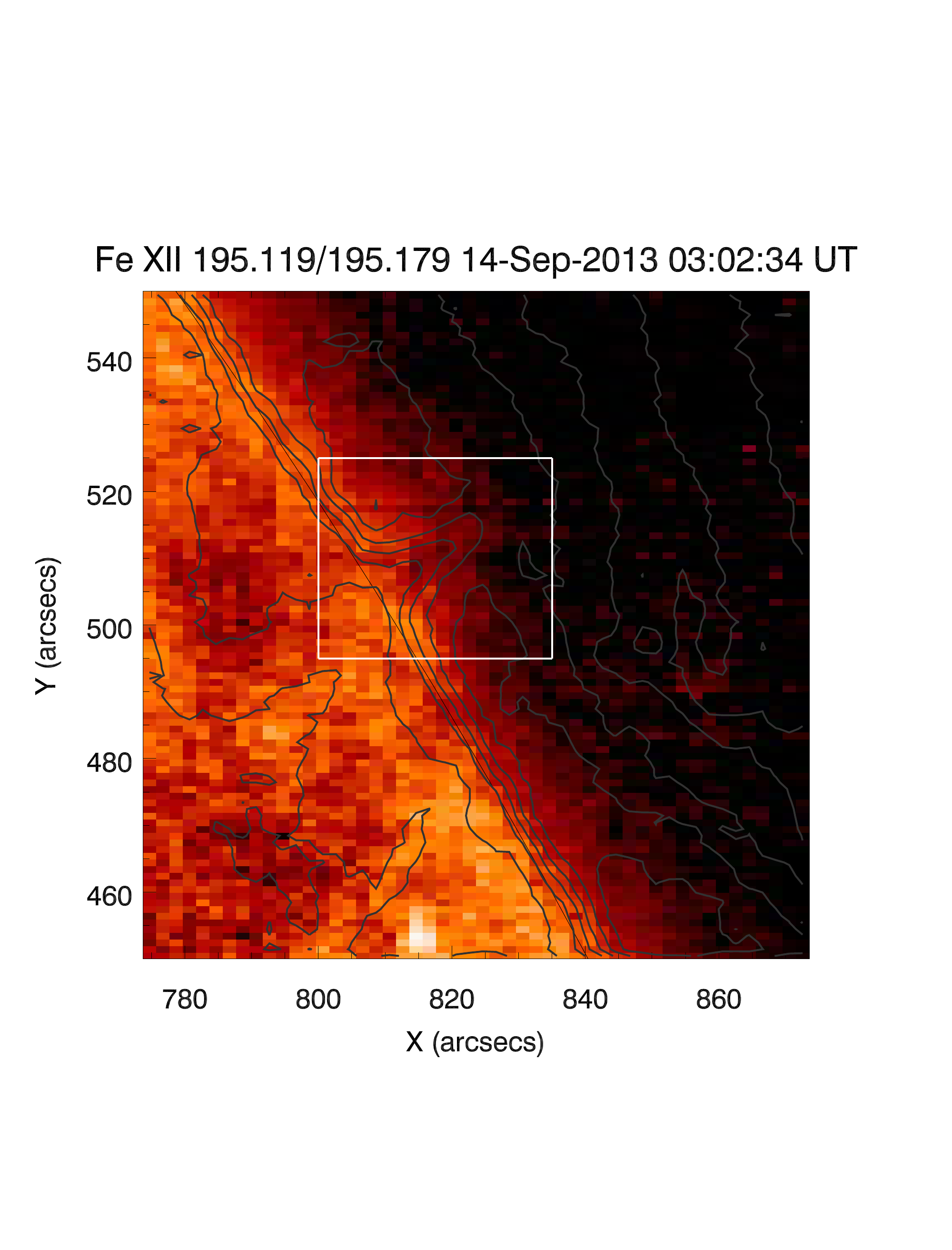}
\includegraphics[scale=0.3]{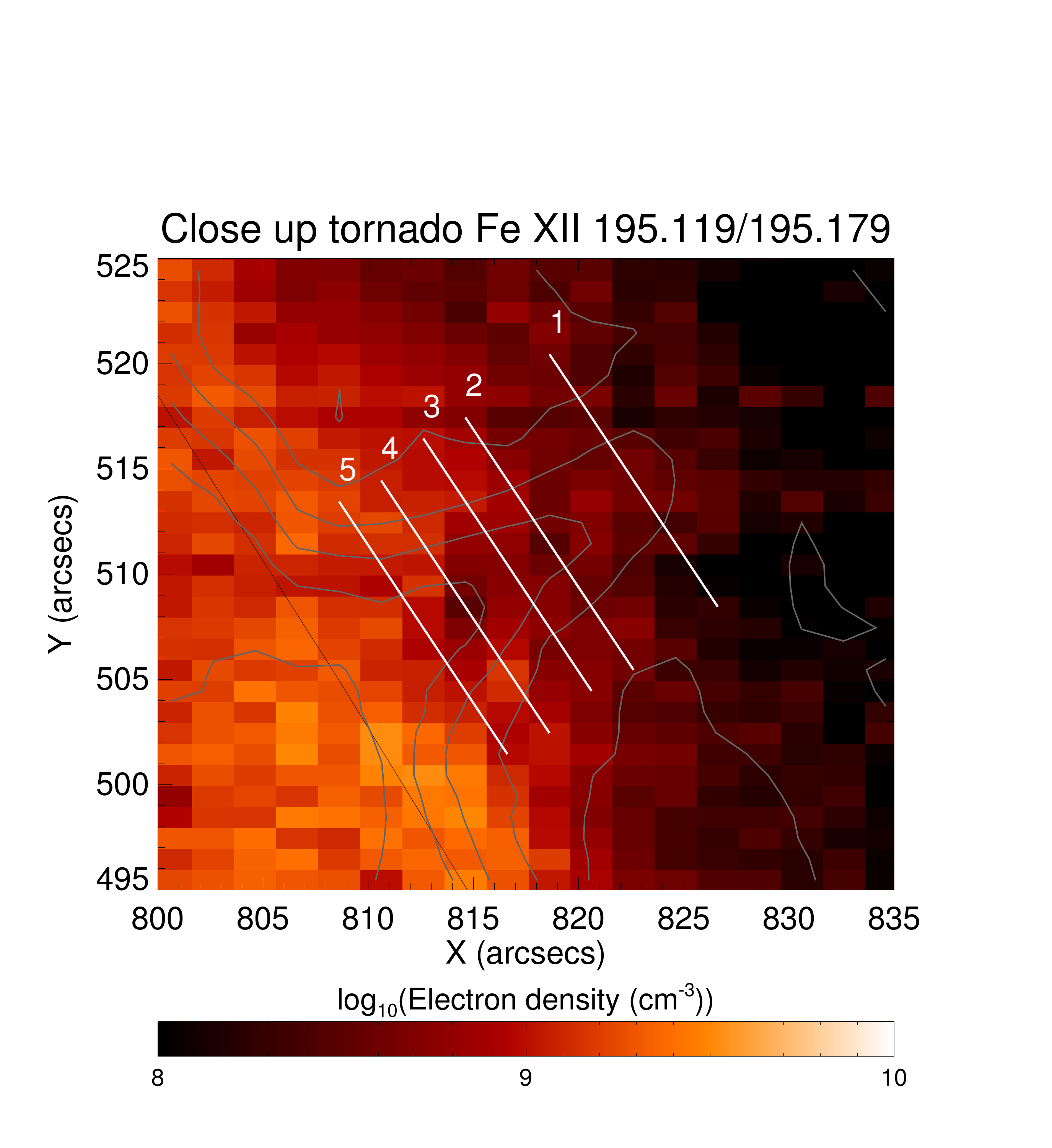}
\caption{Density map for the \ion{Fe}{xii} line pair at 195~\AA. The right panel is a zoomed-in image, the dimensions of which are shown as a white box in the left hand panel. Overplotted on both is the contour of the 195.119~\AA\ line, showing the position of the tornado. Also plotted on the right hand image are five parallel cuts through the tornado axis, used in Figure \ref{fig:195pdfile}}
\label{fig:195dens}
\end{center}
\end{figure*}

\begin{figure}
\begin{center}
\includegraphics[width=\hsize]{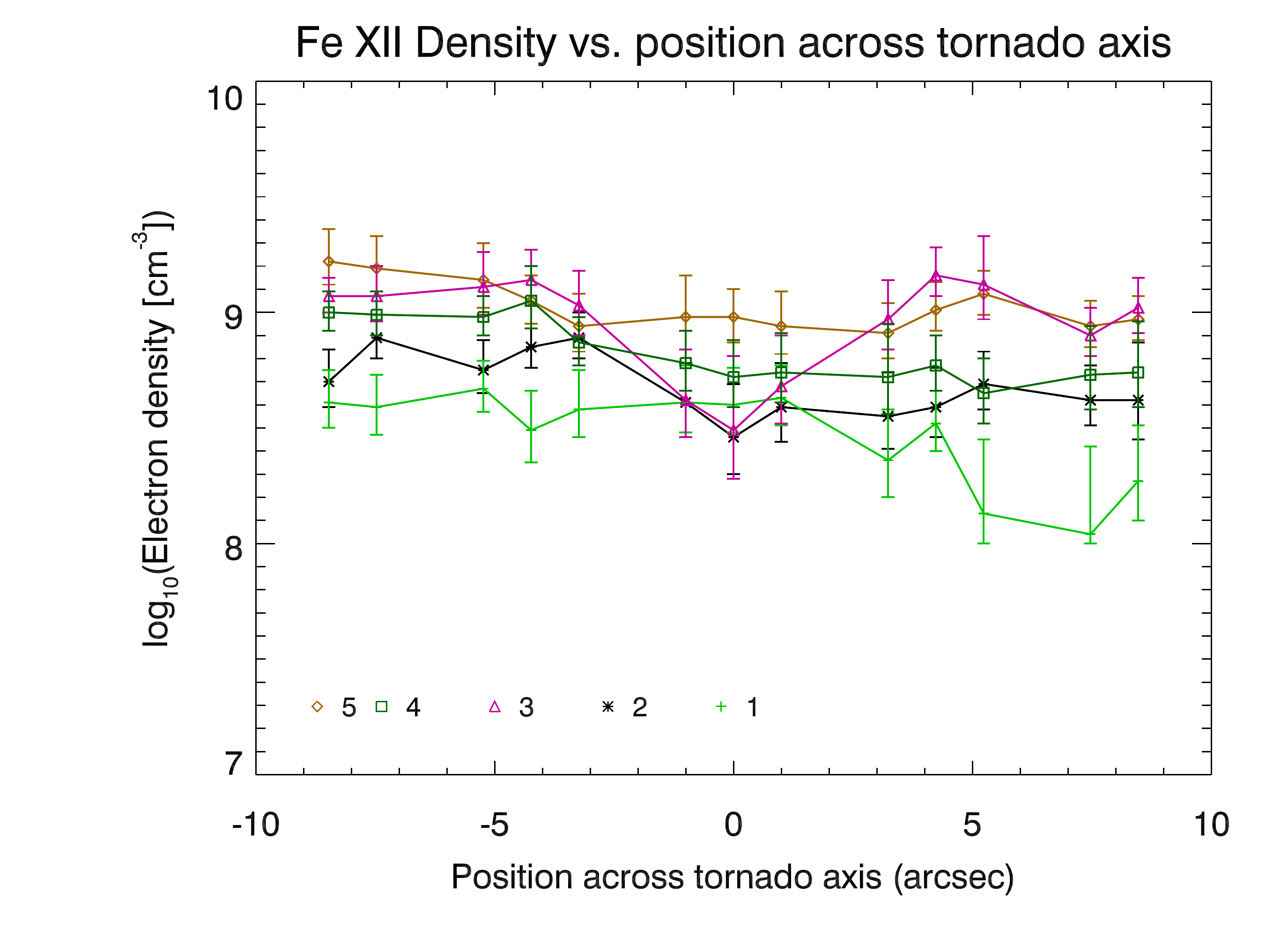}
\caption{Density plotted as a function of position across the tornado axis for five parallel cuts. Cut 1 is the uppermost as seen in Figure \ref{fig:195dens} (farthest from the limb) and cut 5 is closest to the limb. Zero on the x-axis is defined as the centre of the tornado, where the Doppler shift is nearest to zero.}
\label{fig:195pdfile}
\end{center}
\end{figure}

Figure \ref{fig:195dens} shows the {resulting} density map. 
Just like in the \ion{Fe}{xi} maps, we see a small dip in density at the tornado location. This is seen more clearly if we take cuts through the tornado axis, shown as white parallel lines in the right hand panel of Figure \ref{fig:195dens}, the results of which are plotted in Figure \ref{fig:195pdfile}.

Figure \ref{fig:195pdfile} shows how the density changes across the tornado axis. Here we can clearly see that the density is lower towards the middle of the tornado, especially looking at cut 3 (triangle markers, magenta line). This cut reveals a drop in density of around a factor of three when looking towards the centre of the tornado. A number of the other cuts also show a drop in density when looking towards the centre of the tornado, with a minimum at around this position. The uppermost cut (plus markers, light green line) does not display the same pattern, but this location is further into the corona and it is to be expected that the ambient electron density is lower here than nearer to the limb.

\subsection{Density analysis}
\label{ssec:densan}
{From Figure \ref{fig:195pdfile} we observe a mean electron density of $\log{n_e} \sim 8.5 - 9.2$ in the hot ($\log{\mathrm{T}} = 6.2$) plasma. This range is consistent with values found previously \citep[see review by][]{Labrosse10}.}

These observations bring up questions about {the origin of the observed drop in density}. The velocity maps {presented in Figures \ref{fig:197vel} and \ref{fig:185vel}} suggest that the majority of the hot line emission is coming from the tornado -- the strong Doppler structure indicates that these emission lines are formed in the moving plasma. {Hence} we can assume that this dip in density is also due to the tornado itself. We also know that the temperature of this lower density region is around $\log{\mathrm{T}} = 6.2$, as it is visible in all available diagnostics at this temperature. This suggests that the rotating hot plasma is less dense {in the central parts of the tornado than at its edges, while the edges are denser than the surrounding corona.}


\section{Non-thermal line widths}
\label{sec:NTLW}


The measured spectral line width can be attributed to three main factors. The first is the instrumental width, broadening introduced by the optics of the telescope, that has a measured value and can easily be removed. Secondly there is the thermal width, which is associated with the thermal motion due to ambient plasma temperatures. Any remaining line width is the non-thermal line width (NTLW). Assuming that each of these can be characterised by a Gaussian profile, we can split the measured line width, $\Delta \lambda_{meas}$, by equation \ref{eqn:ntlw}:
\begin{equation}
\label{eqn:ntlw}
\Delta \lambda_{meas}^2 = \Delta \lambda_{inst}^2 + \Delta \lambda_{th}^2 + \Delta \lambda_{NT}^2 \ ,
\end{equation}
where $\Delta \lambda_{inst}$ is the instrumental line width, $\Delta \lambda_{th}$ is the thermal width and $\Delta \lambda_{NT}$ is the non-thermal component. The thermal width can be calculated, and the instrumental width of EIS has been measured \citep[see][]{Brown08}.
{Here we adopt the measured laboratory value (0.047~\AA)  for the short wave band CCD as an absolute lower limit for $\Delta \lambda_{inst}$, so that we can ascribe an upper limit to the non-thermal width.}

\begin{figure*}
\begin{center}
\includegraphics[scale=0.4]{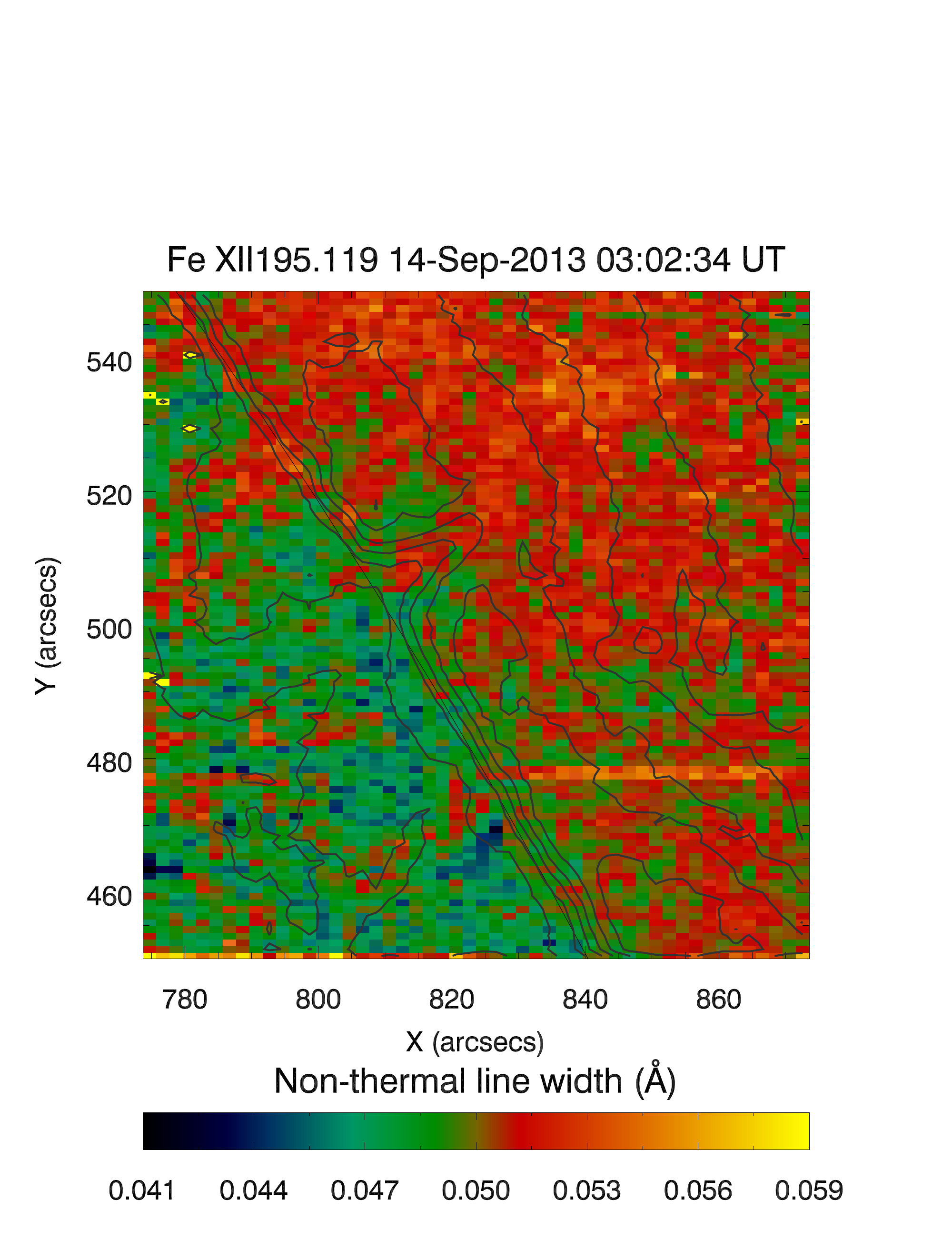}
\includegraphics[scale=0.4]{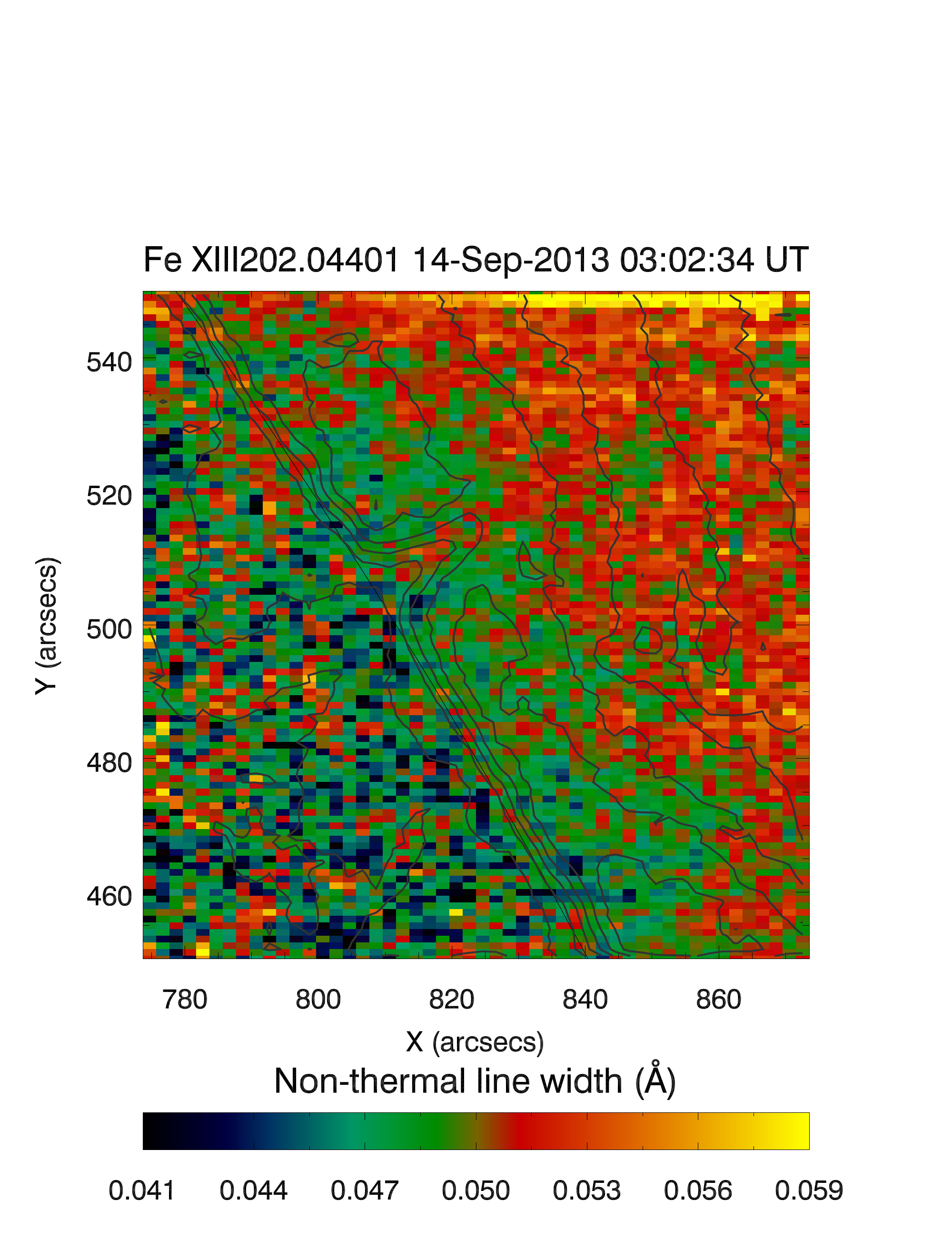}
\caption{Non-thermal line width maps of the \ion{Fe}{xii} 195.119~\AA\ (left) and \ion{Fe}{xiii} 202.044~\AA\ (right) lines. These images show a slightly larger field of view than, for example, Figures \ref{fig:197vel} and \ref{fig:185vel}, but the resolution is the same.}
\label{fig:195wid}
\end{center}
\end{figure*}

With the instrumental and thermal line widths removed from the line profile, we can examine the non-thermal broadening. Figure \ref{fig:195wid} shows the NTLW maps of the \ion{Fe}{xii} 195.119~\AA{, which was de-blended as described in Section \ref{sssec:feXIIblend},} and \ion{Fe}{xiii} 202.044~\AA\ lines {(other lines did not yield clear NTLW maps)}. {Errors were calculated using the errors on the measured line width from the fitting. The errors on the thermal (calculated using CHIANTI values) and instrumental line widths are assumed to be zero.}

\begin{figure}
\begin{center}
\includegraphics[width=\hsize]{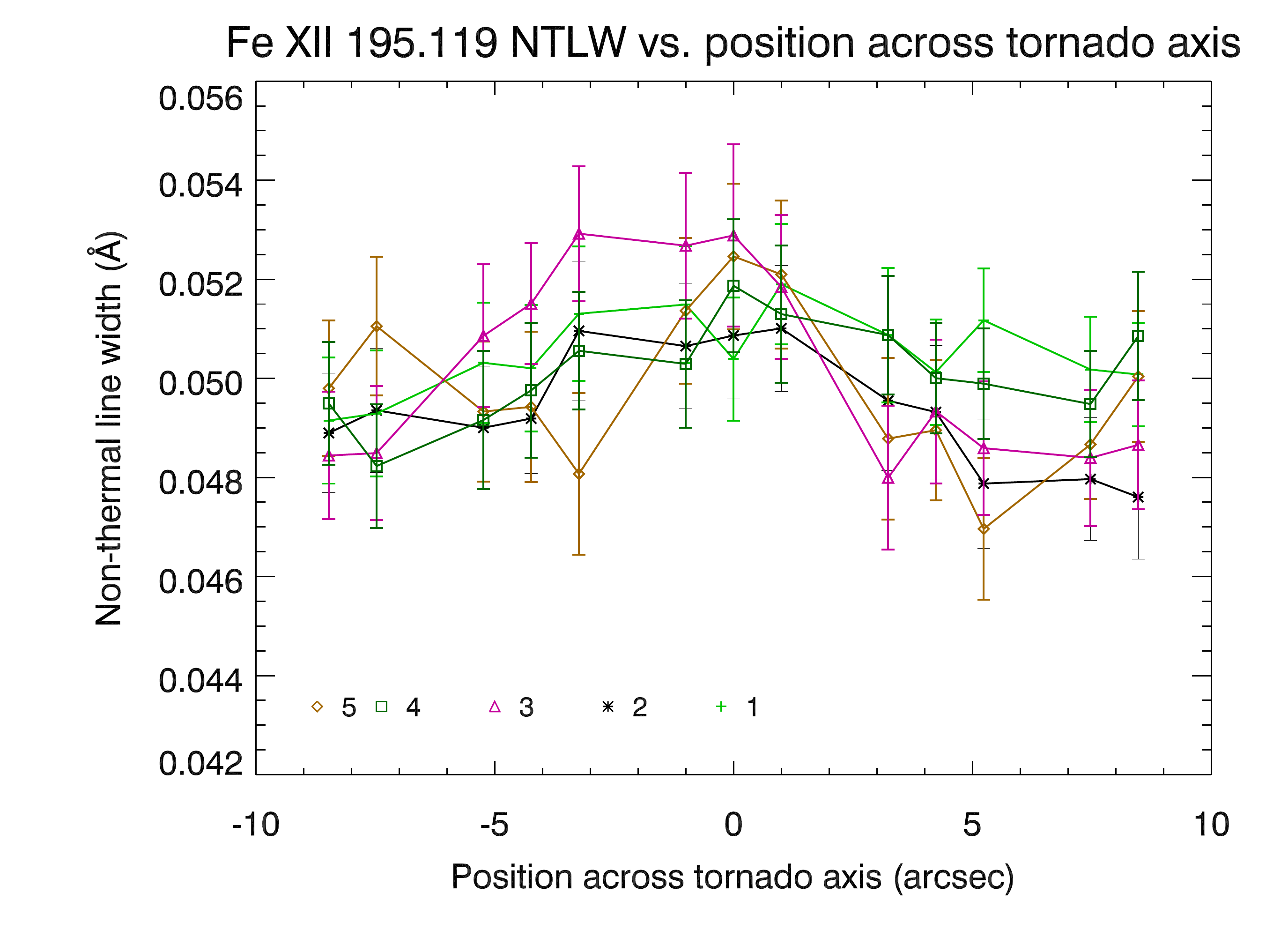}
\caption{{Non-thermal line width plotted as a function of position across the tornado axis for five parallel cuts through the tornado in the NTLW map, here shown for the 195\AA\ \ion{Fe}{xii} line. These are the same cuts used in Figure \ref{fig:195pdfile} and are shown in Figure \ref{fig:195dens}.}}
\label{fig:195widpfile}
\end{center}
\end{figure}
In both cases we see a slightly broadened profile at the tornado location when compared with the corona immediately next to it{, as is emphasised in Figure \ref{fig:195widpfile} which shows cuts through the tornado axis in NTLW. Here we clearly see broader profiles when looking at the tornado}. This would suggest that there is an additional broadening mechanism here.
Although it seems that there is broadening local to the tornado, we cannot take this as conclusive evidence that the tornado is the cause of the broadening. As can be seen in Figure \ref{fig:195wid} there are similar patterns of broadening at other limb locations, so we cannot rule out that this is a chance occurrence.

If we assume, however, that the broadening is indeed caused by the tornado structure, we must consider what mechanisms could be causing this non-thermal broadening. 
{One possibility is related to the possible existence of two types of magneto-hydrodynamic systems in the prominence structure: the densest parts of the prominence plasma would be supported by horizontal and relatively weak magnetic fields, while more dynamic parts of the prominence could be related to strong horizontal fields combined with a turbulent field \citep{2014A&A...569A..85S}. This could explain non-thermal line broadening in the tornado.}
{Alternatively}, it has been suggested (T. Zaqarashvili, {private communication}) that the Kelvin-Helmholtz instability in tornadoes could cause line broadening. This kind of instability occurs when there is a shear flow between two plasmas, much like we would expect at the boundary between a rotating tornado and the surrounding corona. This observation of non-thermal broadening then appears to add weight to the suggestion that the tornado structure is rotating. Observations of the Kelvin-Helmholtz instability have been suggested in relation to observed Rayleigh-Taylor instabilities in quiescent prominences \citep{Berger14}.


\section{Emission measure distribution}
\label{sec:DEM}
{In this section we investigate the emission measure distribution and differential emission measure at four locations in the raster: in the tornado, in the main prominence body, and at two locations in the corona at the same altitude as the tornado and prominence respectively.}

Differential emission measures (DEMs) are a way of identifying the temperature distribution along the line of sight. Using a regularised inversion code from \citet*{Hannah12}, adapted for use of spectral lines available in this EIS data set, we can calculate the regularised DEM at different points across the raster. This code utilises the intensities from the Gaussian fitting procedure described in Section \ref{sec:data} along with contribution functions calculated using CHIANTI \citep{Dere97,Landi12} to perform the inversion \citep[see][for full description of inversion technique]{Hannah12}. {We assumed} photospheric abundances for calculating the DEM {in pixels located in the tornado and prominence body, and coronal abundances for pixels located in the corona}.

DEMs in prominences have been calculated numerous times using a number of different instruments \citep[see, e.g.][]{Schmahl86,Wiik93,Parenti07,Parenti12}, but none have used EIS data for this analysis. {To our knowledge, } no DEMs of solar tornadoes have been published at the time of writing.

\begin{figure*}
\begin{center}
\includegraphics[scale=0.5]{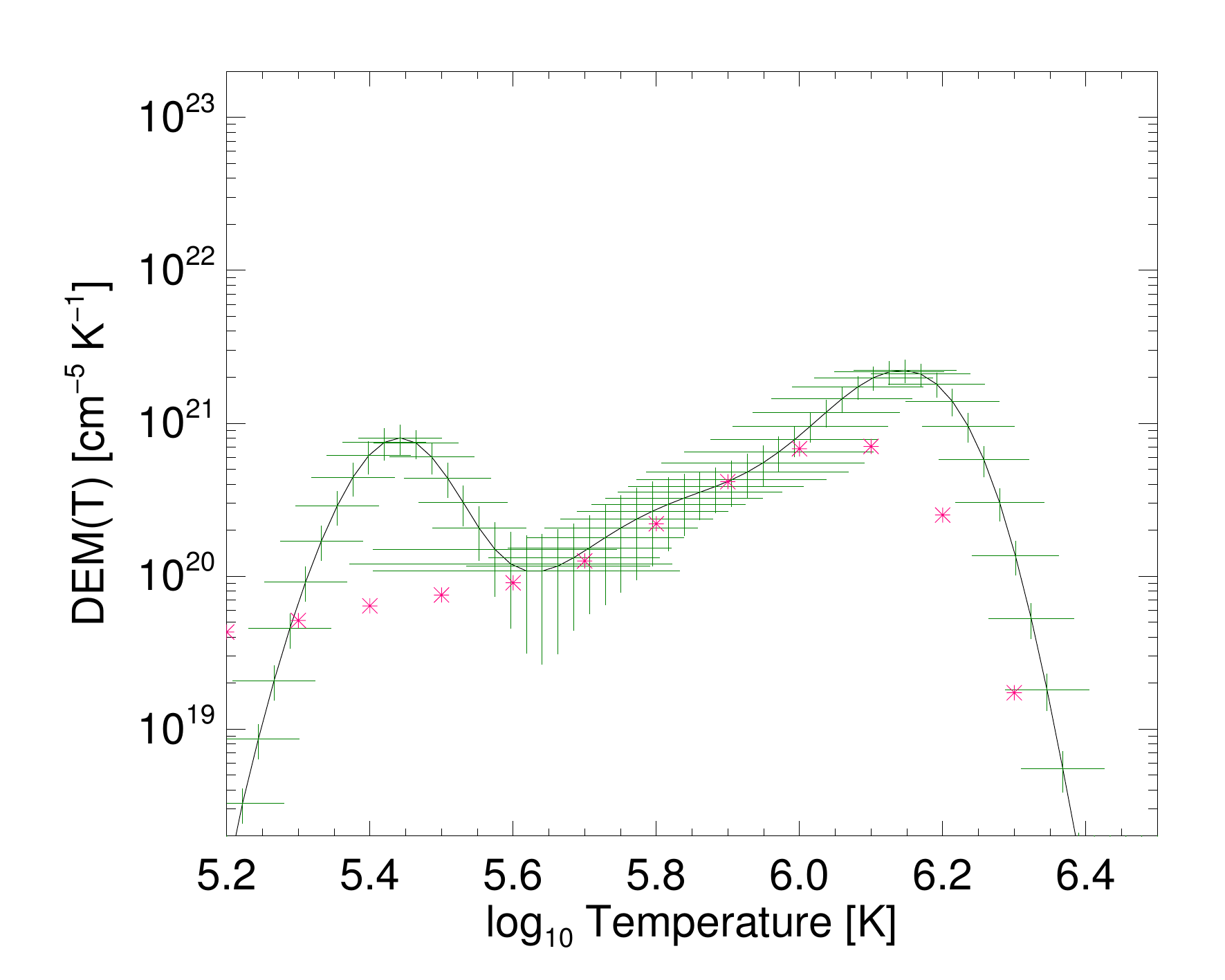}
\includegraphics[scale=0.5]{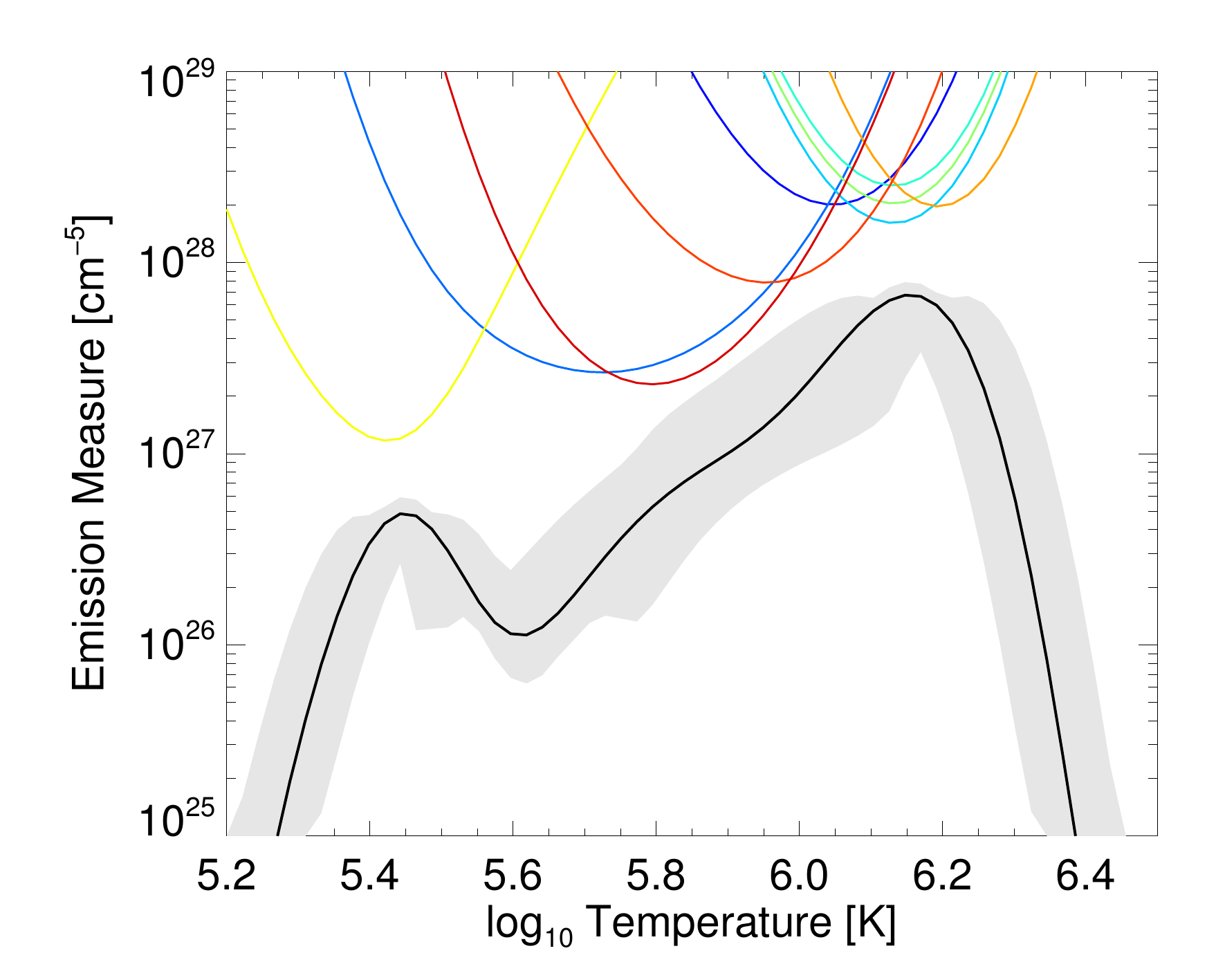}
\caption{Result of the DEM code for a pixel in the tornado (see Figure \ref{fig:195int}). Photospheric abundances were assumed for this DEM. Plotted here is the resulting differential emission measure (left panel) along with the emission measure distribution (right panel). Also plotted on the EMD panel {(right)} are the EM loci curves, or constraint curves, for each of the lines used in the DEM calculation {(see Table \ref{tab:DEMlines} for line identification and formation temperatures)}. The green `error bars' in the {DEM} (left panel) and the grey ranges in the {EMD} (right panel) represent confidence regions for the DEM and EMD fits respectively. {Plotted as pink asterisk markers on the DEM are the results from \citet{Parenti07} for a quiescent prominence.}}
\label{fig:DEMtornado}
\end{center}
\end{figure*}


\begin{figure*}
\begin{center}
\includegraphics[scale=0.5]{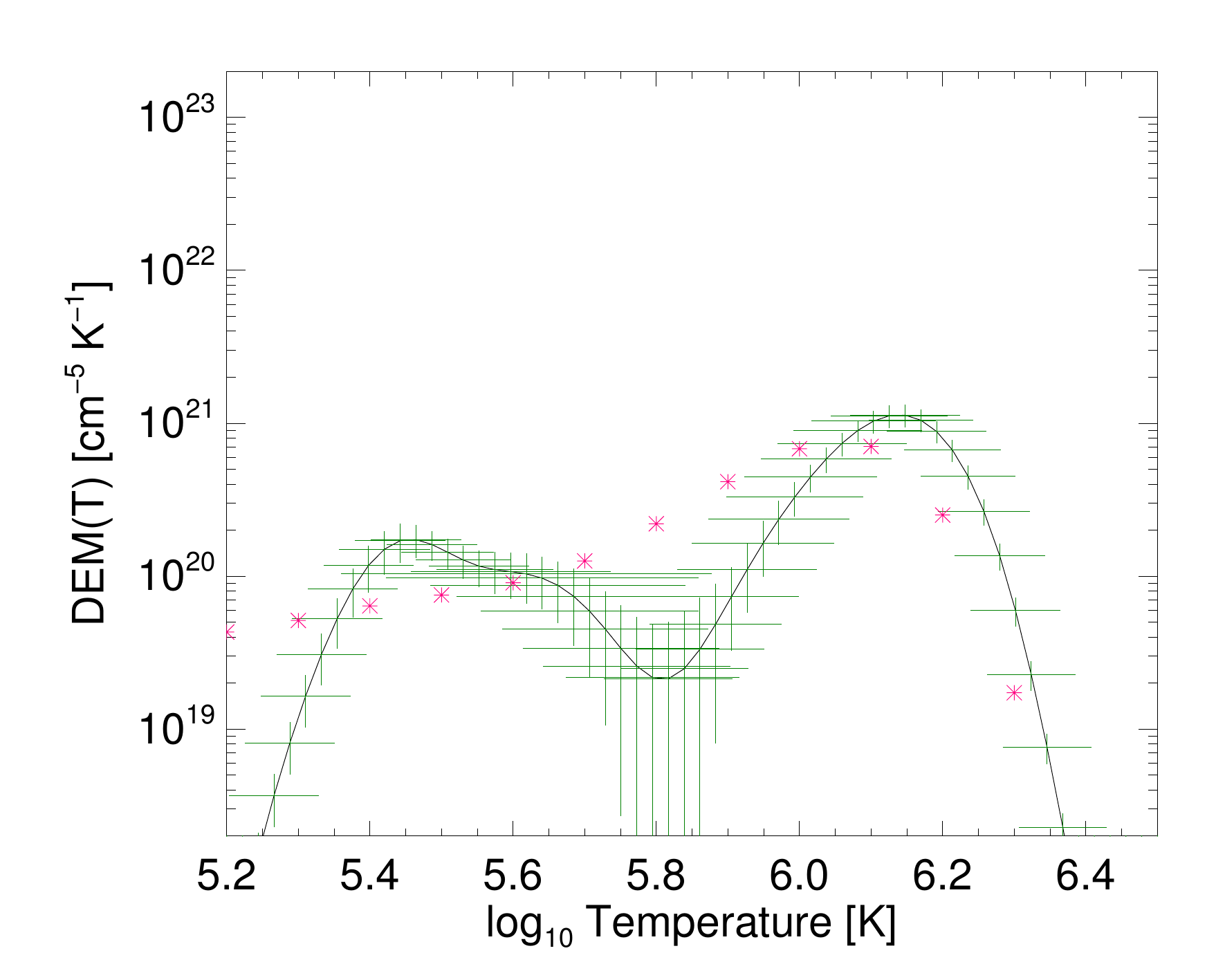}
\includegraphics[scale=0.5]{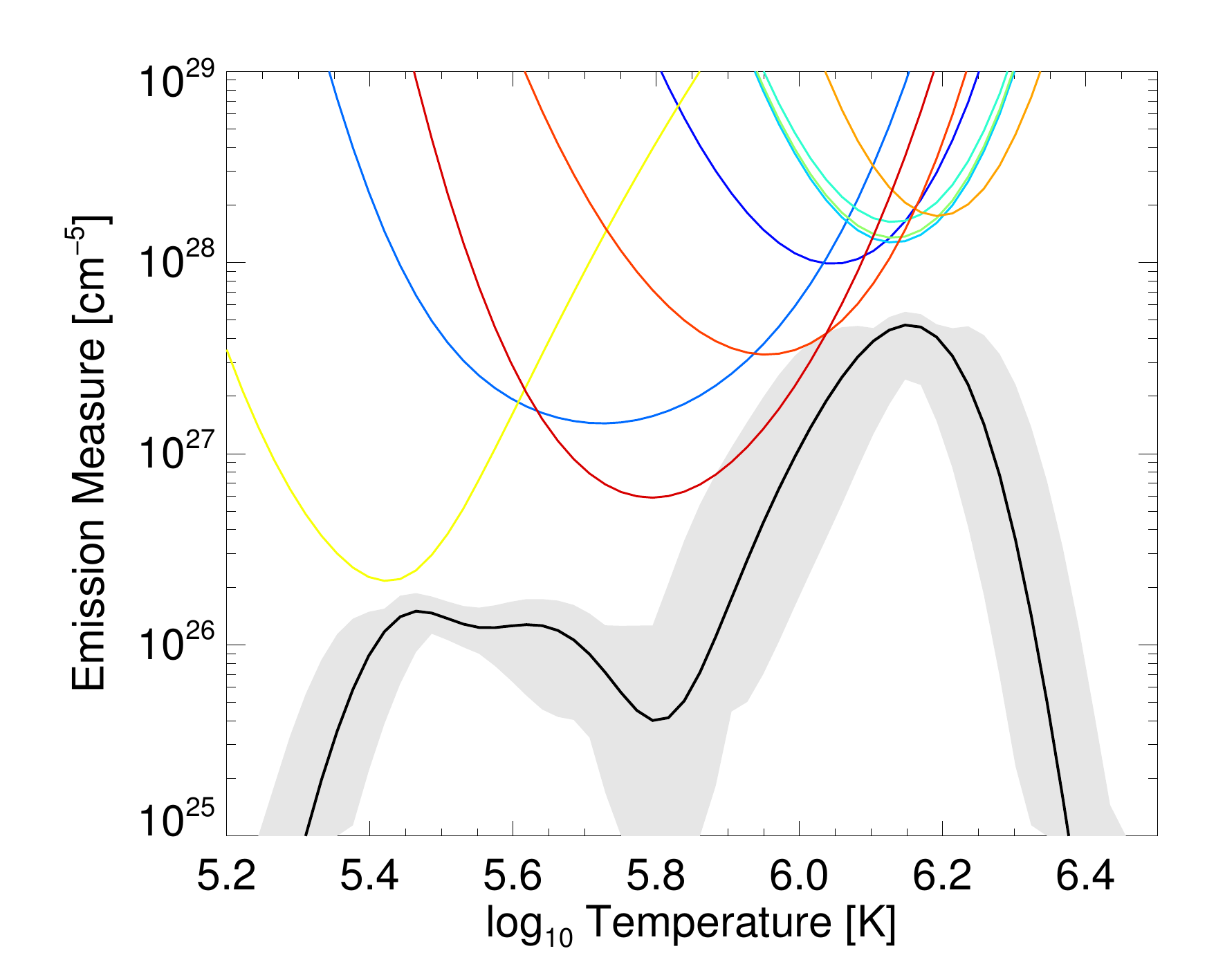}
\caption{{Same as Fig.~\ref{fig:DEMtornado} for a pixel in the main prominence body}.}
\label{fig:DEMprom}
\end{center}
\end{figure*}


\begin{figure*}
\begin{center}
\includegraphics[scale=0.5]{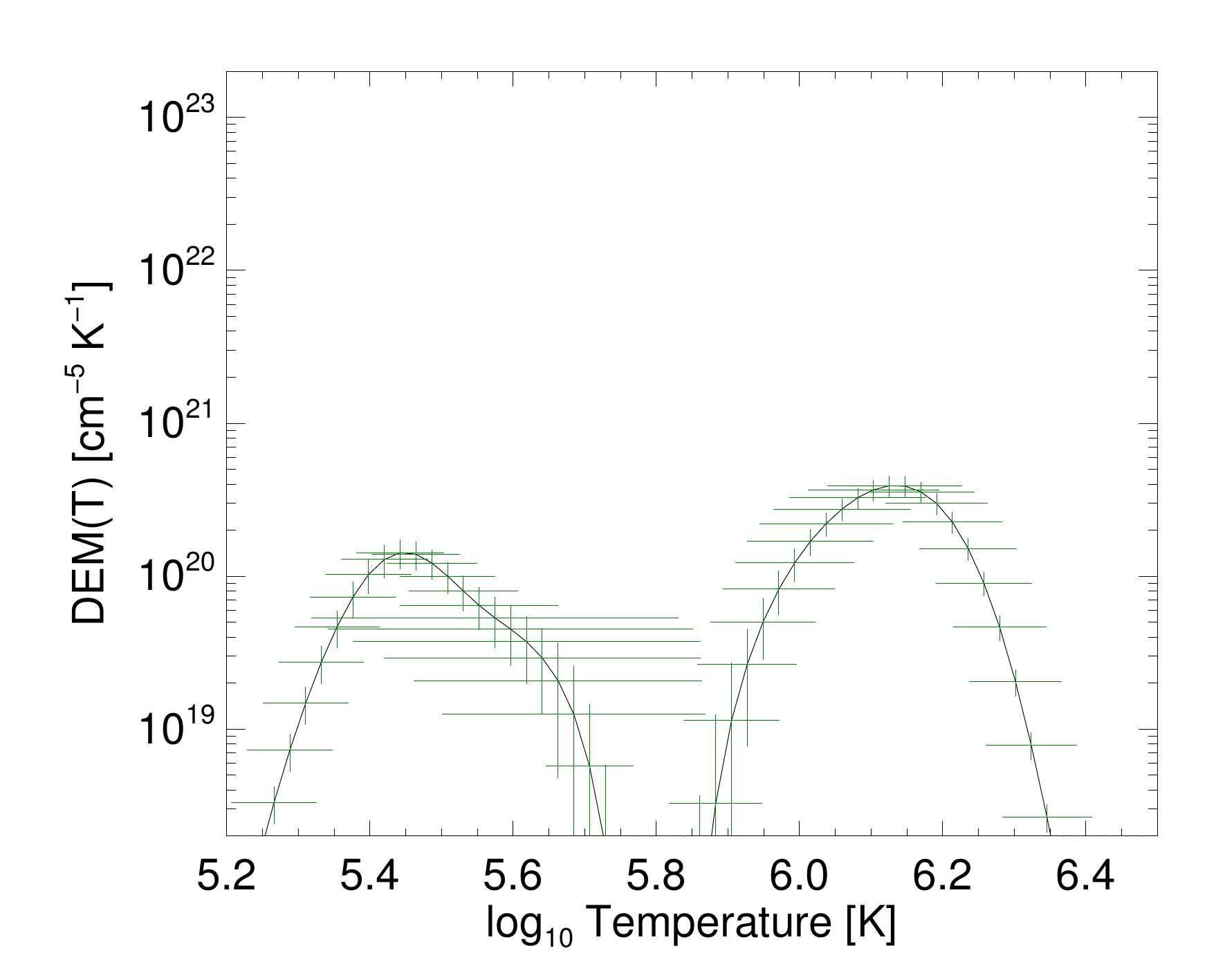}
\includegraphics[scale=0.5]{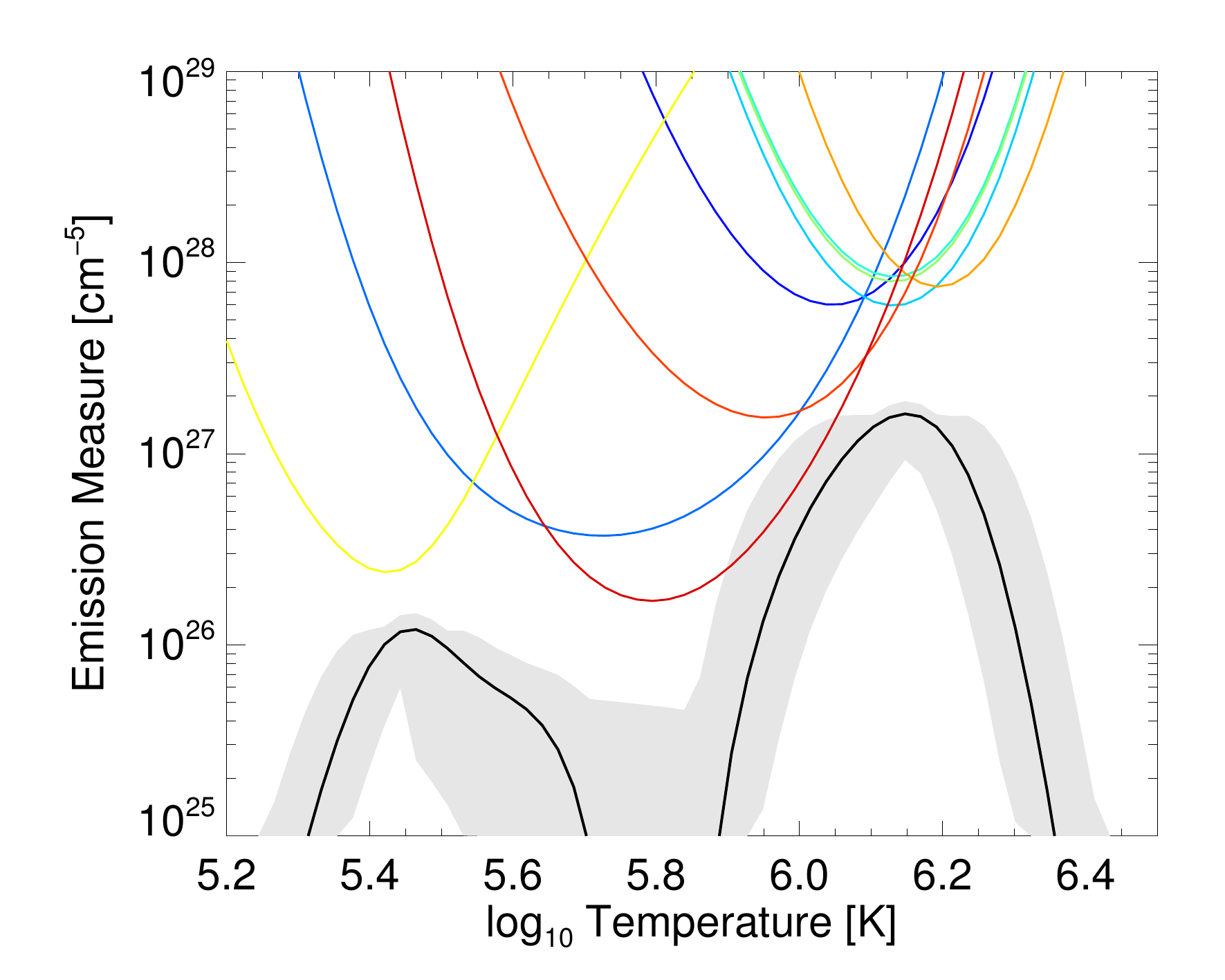}
\caption{{DEM (left) and EMD (right) for a point in the corona at {a similar} altitude as the tornado DEM (Figure \ref{fig:DEMtornado}).}}
\label{fig:DEMcorona2}
\end{center}
\end{figure*}



\begin{figure*}
\begin{center}
\includegraphics[scale=0.5]{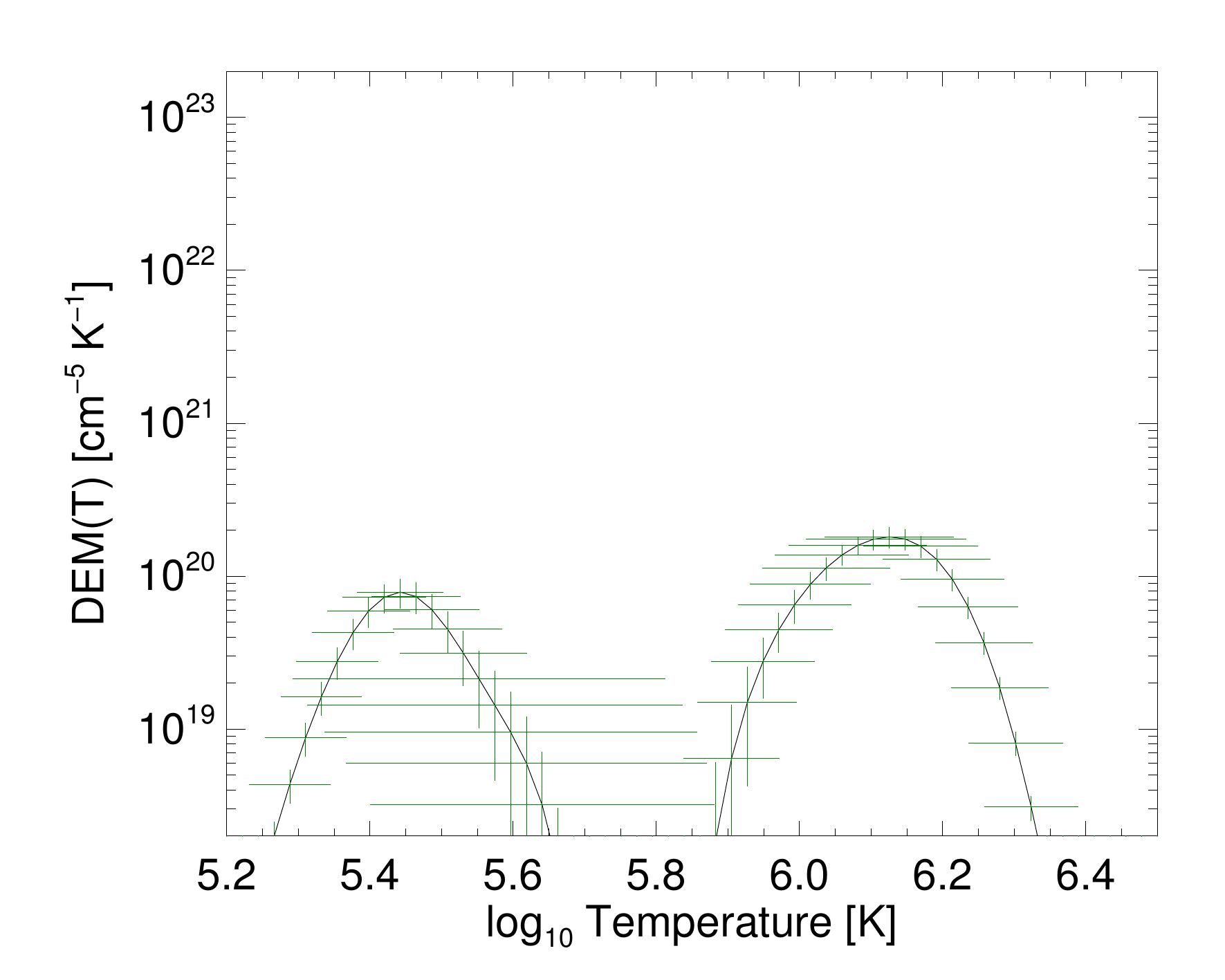}
\includegraphics[scale=0.5]{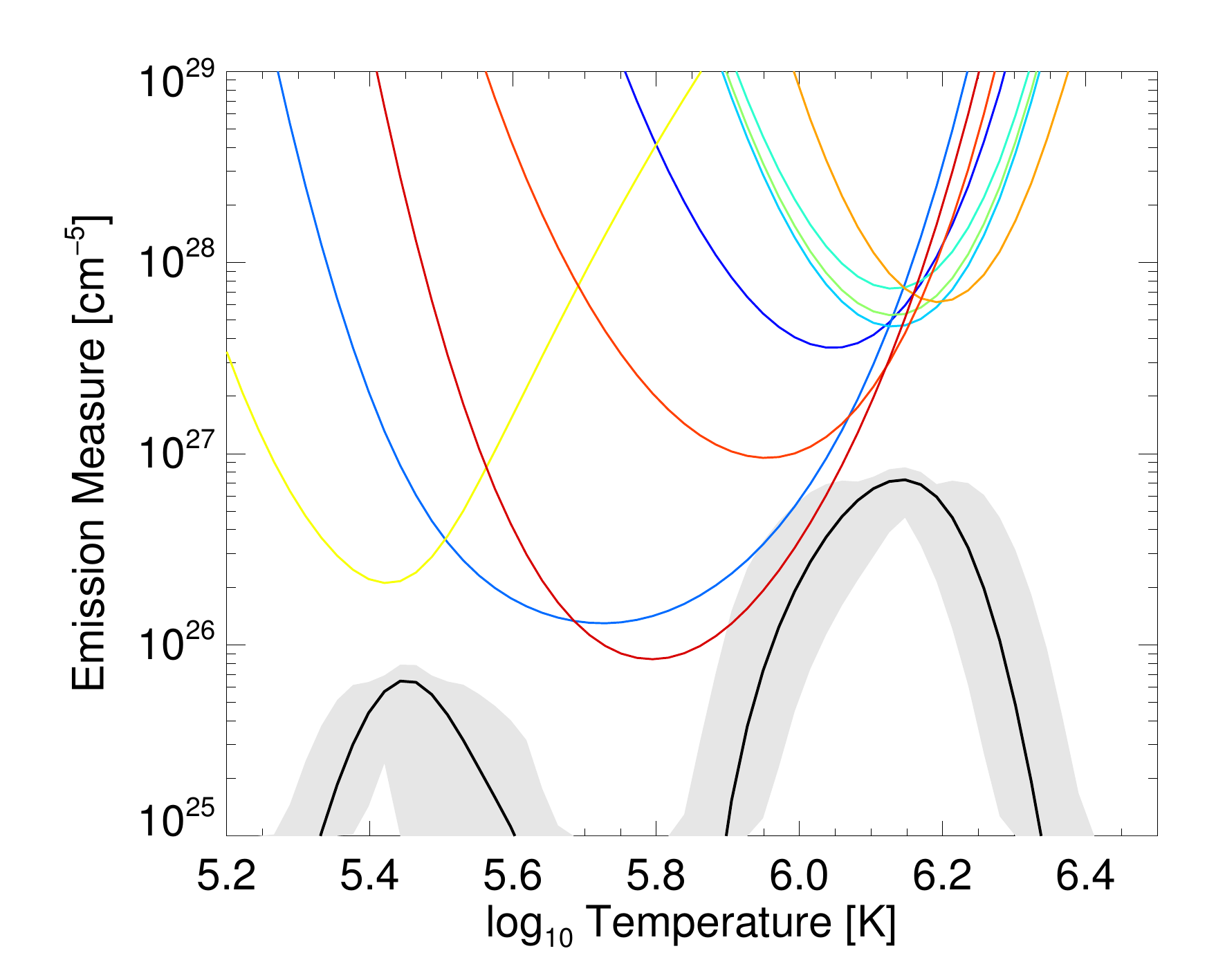}
\caption{{DEM (left) and EMD (right) for an arbitrary point in the corona at a similar height above the limb to the prominence body DEM of Figure \ref{fig:DEMprom}.}}
\label{fig:DEMcorona}
\end{center}
\end{figure*}

Table \ref{tab:DEMlines} contains information on the EIS lines that have been used when calculating these DEMs. 
\begin{table}
\caption{List of lines available in the EIS study for constraining the DEM.}
\label{tab:DEMlines}
\centering
\begin{tabular}{c c c}
\hline\hline
Ion & $\lambda_0$ (\AA) & $\log{\mathrm{T (K)}}$\\
\hline
\ion{O}{v} & 192.904 & 5.4\\
\ion{Fe}{viii} & 185.213 & 5.7\\
\ion{Si}{vii} & 275.361 & 5.8\\
\ion{Fe}{ix} & 197.862 & 6.0\\
\ion{Fe}{x} & 184.537 & 6.1\\
\ion{Fe}{xi} & 188.216 & 6.2\\
\ion{Fe}{xi} & 188.299 & 6.2\\
\ion{Fe}{xi} & 192.814 & 6.2\\
\ion{Fe}{xii} & 195.119 & 6.2\\
\hline
\end{tabular}
\end{table}
Some lines have been omitted from the DEM for a number of reasons. {For example, }the coolest line in the study, \ion{He}{ii} at 256~\AA, must be ignored due to the fact that it is optically thick, as well as being heavily blended. \ion{Fe}{xiii} 202.044~\AA, the hottest line available in the raster, was also not used as it is found to be density sensitive at densities below around $\log{n_e} = 10$ \citep[see figure 4.2 in][]{Davethesis}. In total, we are left with nine lines that can be used for the DEM.

{For this analysis errors of 22\% have been assumed for the line intensities. This is considerably higher than the errors from line fitting, but due to combined uncertainties from the response function calibration of the EIS instrument \citep{Lang06} we assume here the larger percentage error. This calibration uncertainty was measured pre-flight, and so must be considered as a lower limit.}

Figure {\ref{fig:DEMtornado} {(left panel)} shows a DEM produced from a pixel in the tornado, and Figure} \ref{fig:DEMprom} {(left panel)} shows {that} from a pixel in the main prominence body {(indicated by plus signs in Fig.~\ref{fig:195int})}. {We recover a different solution for each of these regions, with the tornado (Figure \ref{fig:DEMtornado}) showing more plasma at $\log{\mathrm{T}} = 5.4$ than in the prominence: the DEM is larger in the tornado than {in} the prominence by a factor of {4 -- 5} at this temperature. This suggests that there is more cool material along the line of sight towards the tornado, which is {supported by the fact that when observed in coronal lines, the tornado appears darker than the prominence, signalling greater absorption of the background emission.}}

{Figures \ref{fig:DEMcorona2} and \ref{fig:DEMcorona} show DEM and EMD plots for pixels in the corona, at similar altitudes to the DEMs presented in Figures \ref{fig:DEMtornado} and \ref{fig:DEMprom} respectively. We here present these coronal DEMs to show that it is not just the foreground corona that we are seeing in the prominence and tornado DEMs. Figure \ref{fig:DEMcorona2} {(for a pixel in the corona, close to the tornado)} contains one main peak{, with $\mathrm{DEM(T)} = 4 \times 10^{20}$ cm$^{-5}$ K$^{-1}$,} at around $\log{\mathrm{T}} = 6.2$, and a smaller {($\mathrm{DEM(T)} \sim 1.5 \times 10^{20}$ cm$^{-5}$ K$^{-1}$)}, less well constrained contribution at around $\log{\mathrm{T}} = 5.5$. {In contrast, the tornado DEM (Figure \ref{fig:DEMtornado}) has a lower temperature peak at $\log{\mathrm{T}} = 5.4$, where $\mathrm{DEM(T)} = 8 \times 10^{20}$,}
~suggesting that there is {much} more contribution at lower temperatures coming from the tornado itself.
~In Figure \ref{fig:DEMcorona} {(for a pixel in the corona at the same altitude as the prominence body)} we see {little} contribution below $\log{\mathrm{T}} = 6.0${, consistent with the findings of previous authors \citep[e.g.][]{Landi10}}. {The coronal DEMs (Figures \ref{fig:DEMcorona2} and \ref{fig:DEMcorona}) confirm that the peaks around $\log{\mathrm{T}} = 6.2$ observed in the DEMs of the tornado and the prominence (Figures \ref{fig:DEMtornado} and \ref{fig:DEMprom}) could be due to foreground emission.}

Taking an estimate of the gradient of the slope of these DEM curves between $\log{\mathrm{T}} = 5.65$ and $\log{\mathrm{T}} = 6.10$ for the tornado {and $\log{\mathrm{T}} = 5.88$ and $\log{\mathrm{T}} = 6.10$ for the prominence body}, we recover a value of {$2.6^{+0.7}_{-0.5}$} for the tornado (Figure \ref{fig:DEMtornado}) and {{$6.4^{+4.3}_{-1.7}$} for the prominence body (Figure \ref{fig:DEMprom})}.

The coloured lines in the {right} panels of Figures \ref{fig:DEMtornado}, \ref{fig:DEMprom},~\ref{fig:DEMcorona2} {and \ref{fig:DEMcorona}} represent the constraint curves for the EMDs. Even though we have a limited number of lines, {there is} still quite a good temperature {range available to constrain the curves}. More lines formed at cooler temperatures would have helped, as prominences generally consist of lower temperature plasma.

{Plotted as asterisks in the DEM panels of Figures \ref{fig:DEMtornado} and \ref{fig:DEMprom} are the results of \citet{Parenti07} for a quiescent prominence. In that paper they used \textit{SOHO}/SUMER data on a large prominence, finding a high temperature peak in the DEM at around $\log{\mathrm{T}} = 6.0$, by using a selection of lines from the SUMER spectrum. The analysis done here is on a much smaller prominence, only 20\arcsec\ in altitude, using a smaller number of EIS lines, most of which are formed at higher temperatures than those used by \citeauthor{Parenti07}. {Those authors also used an older version of the CHIANTI database for their analysis (version 4.2, as opposed to version 7.1 used here, which has updated Fe ionisation temperatures).} The fact that we find a peak at $\log{\mathrm{T}} = 6.2$ is therefore not surprising.} {Our prominence is at a lower altitude, which } {means that there is more hot coronal emission along the line of sight than in the \citeauthor{Parenti07} case. Also, the selection of EIS lines used means that} {our} {DEM is better constrained at this temperature than at lower temperatures, where there are limited lines available. The coolest line available for this DEM was \ion{O}{v} at $\log{\mathrm{T}} = 5.4$. Below this temperature there are no constraints, and we see a sharp drop-off that therefore cannot be associated with any observational effects.}

{We note here that in the range $\log{\mathrm{T}} = 5.6 - 6.0$ the slope of the tornado DEM {(Figure \ref{fig:DEMtornado})} matches that of the \citeauthor{Parenti07}} {DEM}.


\section{Discussions and Conclusions}
\label{sec:conc}
{This paper  presents a number of plasma diagnostics using an EIS data set that captured a tornado above the solar limb}. We find a similar line-of-sight velocity pattern to \citet{Su14} {at $\log{\mathrm{T}} = 6.2$. We show that this pattern in the line-of-sight velocities, suggestive of rotational motion, persists at temperatures down to $\log{\mathrm{T}} = 6.0$ with the \ion{Fe}{ix} 197.862~\AA\ line, and up to temperatures of $\log{\mathrm{T}} = 6.3$, with the \ion{Fe}{xiii} line at 202.044~\AA.} Lines formed at temperatures lower than $\log{\mathrm{T}} = 6.0$ do not show any signs of a Doppler split about the tornado axis. It should be noted, however, that it may still be the case that this pattern exists at lower plasma temperatures, but for reasons discussed in Section \ref{sec:vel}, we cannot see them with this data set.

Using the CHIANTI atomic database, we have  been able to give an estimate of {the} electron density in the region of the tornado {based on \ion{Fe}{xi} and \ion{Fe}{xii} lines, both of which are formed at around $\log{\mathrm{T}} = 6.2$}. {At this temperature we obtain an electron density of $\log{n_e} = 8.5$ when looking directly towards the centre of the tornado. In the surrounding plasma, however, we recover $\log{n_e} \sim 9$. In other words, in} all diagnostics available we see a dip in density when looking towards the tornado location, compared with the ambient corona. 

{These results} suggest that the  hot moving plasma  forms a sheath region of lower density around the cool core. {On the edge of the tornado, we could be seeing denser, cooling threads. This picture seems consistent with the multi-threaded prominence model of \cite{2012ApJ...746...30L}.}

The non-thermal line widths {inferred from our analysis} {may indicate} broader line profiles for two Fe lines at the tornado location, though this result is not entirely certain. If this broadening is, however, localised to and caused by the tornado, then {this could be explained by the presence of turbulence, or  by the Kelvin-Helmholtz instability, or possibly other mechanisms}.

{We  provide the first prominence DEMs using EIS data. {Previous  authors, using other instruments, } found that prominence DEMs peak at around $\log{\mathrm{T}} = 6.0$ \citep{Wiik93,Parenti07}, whereas  we find a peak at $\log{\mathrm{T}} = 6.2$. We attribute this difference to the selection of lines available from EIS, which are mostly formed at hot coronal temperatures, {differences in the atomic data used,} as well as differences in the relative altitudes of the prominences in question.
We also find differences between the DEM measured in the tornado and that from the main prominence body. In the tornado there appears to be more contribution at lower temperatures ($\log{\mathrm{T}} = 5.4$) than we find in the prominence. 
We also find differences in the gradients of the {tornado} DEM{, taken} between $\log{\mathrm{T}} = 5.65$ and $\log{\mathrm{T}} = 6.10$, { and the prominence body DEM, taken between $\log{\mathrm{T}} = 5.88$ and $\log{\mathrm{T}} = 6.10$.} In the tornado we have a gradient of $2.6^{+0.7}_{-0.5}$, {and in the prominence body it is $6.4^{+4.3}_{-1.7}$.}}

Following this work we aim to go on to investigate the magnetic structure of these tornado features, as well as the plasma {properties over a broader temperature range}. Along with the {present study}, these should help build a more accurate picture of the nature of solar tornadoes and how they relate to the overall prominence structure.


\begin{acknowledgements}
P.J.L. acknowledges support from an STFC Research Studentship. N.L. and L.F. acknowledge support from STFC grants ST/I001808/1 and ST/L000741/1. \textit{Hinode} is a Japanese mission developed and launched by ISAS/JAXA, with NAOJ as domestic partner and NASA and STFC (UK) as international partners. It is operated by these agencies in co-operation with ESA and NSC (Norway). CHIANTI is a collaborative project involving George Mason University, the University of Michigan (USA) and the University of Cambridge (UK). SMART is operated by Kwasan and Hida Observatories, Graduate School of Science, Kyoto University, Japan. The AIA data are provided courtesy of NASA/\textit{SDO} and the AIA science team.
{The authors thank S. Parenti for providing the data for the DEM published in \cite{Parenti07}, and I. Hannah for providing valuable information on analysing the DEMs provided by the \cite{Hannah12} code and the use of uncertainties in this code.}
\end{acknowledgements}


\bibliographystyle{aa}
\bibliography{bibliography}

\begin{thebibliography}{32}
\expandafter\ifx\csname natexlab\endcsname\relax\def\natexlab#1{#1}\fi

\bibitem[{Anzer \& Heinzel(2005)}]{Anzer05}
Anzer, U. \& Heinzel, P. 2005, \apj, 622, 714

\bibitem[{Berger(2014)}]{Berger14}
Berger, T. 2014, IAU Symposium, 300, 15

\bibitem[{Brown {et~al.}(2008)Brown, Feldman, Seely, Korendyke, \&
  Hara}]{Brown08}
Brown, C., Feldman, U., Seely, J., Korendyke, C., \& Hara, H. 2008, ApJ Suppl.
  Ser., 176, 511

\bibitem[{{Culhane} {et~al.}(2007){Culhane}, {Harra}, {James}, {Al-Janabi},
  {Bradley}, {Chaudry}, {Rees}, {Tandy}, {Thomas}, {Whillock}, {Winter},
  {Doschek}, {Korendyke}, {Brown}, {Myers}, {Mariska}, {Seely}, {Lang}, {Kent},
  {Shaughnessy}, {Young}, {Simnett}, {Castelli}, {Mahmoud}, {Mapson-Menard},
  {Probyn}, {Thomas}, {Davila}, {Dere}, {Windt}, {Shea}, {Hagood}, {Moye},
  {Hara}, {Watanabe}, {Matsuzaki}, {Kosugi}, {Hansteen}, \&
  {Wikstol}}]{Culhane07}
{Culhane}, J.~L., {Harra}, L.~K., {James}, A.~M., {et~al.} 2007, Sol. Phys.,
  243, 19

\bibitem[{{Del Zanna}(2013)}]{delZanna13}
{Del Zanna}, G. 2013, \aap, 555, A47

\bibitem[{Dere {et~al.}(1997)Dere, Landi, Mason, Monsignori~Fossi, \&
  Young}]{Dere97}
Dere, K., Landi, E., Mason, H., Monsignori~Fossi, B., \& Young, P. 1997, A+A
  Suppl. Ser., 125, 149

\bibitem[{Graham(2014)}]{Davethesis}
Graham, D. 2014, PhD thesis, University of Glasgow

\bibitem[{{Graham} {et~al.}(2013){Graham}, {Hannah}, {Fletcher}, \&
  {Milligan}}]{Graham13}
{Graham}, D.~R., {Hannah}, I.~G., {Fletcher}, L., \& {Milligan}, R.~O. 2013,
  \apj, 767, 83

\bibitem[{Hannah \& Kontar(2012)}]{Hannah12}
Hannah, I. \& Kontar, E. 2012, A+A, 539, A146

\bibitem[{{Haugan}(1999)}]{Haugan99}
{Haugan}, S.~V.~H. 1999, \solphys, 185, 275

\bibitem[{{Heinzel} {et~al.}(2008){Heinzel}, {Schmieder}, {F{\'a}rn{\'{\i}}k},
  {Schwartz}, {Labrosse}, {Kotr{\v c}}, {Anzer}, {Molodij}, {Berlicki},
  {DeLuca}, {Golub}, {Watanabe}, \& {Berger}}]{2008ApJ...686.1383H}
{Heinzel}, P., {Schmieder}, B., {F{\'a}rn{\'{\i}}k}, F., {et~al.} 2008, \apj,
  686, 1383

\bibitem[{{Ko} {et~al.}(2009){Ko}, {Doschek}, {Warren}, \& {Young}}]{Ko09}
{Ko}, Y.-K., {Doschek}, G.~A., {Warren}, H.~P., \& {Young}, P.~R. 2009, \apj,
  697, 1956

\bibitem[{Labrosse {et~al.}(2010)Labrosse, Heinzel, Vial, Kucera, Parenti,
  Gun\'{a}r, Schmieder, \& Kilper}]{Labrosse10}
Labrosse, N., Heinzel, P., Vial, J.-C., {et~al.} 2010, Space Sci. Rev., 151,
  243

\bibitem[{Labrosse {et~al.}(2011)Labrosse, Schmieder, Heinzel, \&
  Watanabe}]{Labrosse11}
Labrosse, N., Schmieder, B., Heinzel, P., \& Watanabe, T. 2011, A+A, 531, A69

\bibitem[{Landi {et~al.}(2012)Landi, Del~Zanna, Young, Dere, \&
  Mason}]{Landi12}
Landi, E., Del~Zanna, G., Young, P., Dere, K., \& Mason, H. 2012, A+A Suppl.
  Ser., 744, 99

\bibitem[{{Landi} \& {Young}(2010)}]{Landi10}
{Landi}, E. \& {Young}, P.~R. 2010, \apj, 714, 636

\bibitem[{{Lang} {et~al.}(2006){Lang}, {Kent}, {Paustian}, {Brown}, {Keyser},
  {Anderson}, {Case}, {Chaudry}, {James}, {Korendyke}, {Pike}, {Probyn},
  {Rippington}, {Seely}, {Tandy}, \& {Whillock}}]{Lang06}
{Lang}, J., {Kent}, B.~J., {Paustian}, W., {et~al.} 2006, \ao, 45, 8689

\bibitem[{Li {et~al.}(2012)Li, Morgan, Leonard, \& Jeska}]{Li12}
Li, X., Morgan, H., Leonard, D., \& Jeska, L. 2012, ApJ Letters, 752, L22

\bibitem[{{Luna} {et~al.}(2012){Luna}, {Karpen}, \&
  {DeVore}}]{2012ApJ...746...30L}
{Luna}, M., {Karpen}, J.~T., \& {DeVore}, C.~R. 2012, \apj, 746, 30

\bibitem[{{Orozco Su{\'a}rez} {et~al.}(2012){Orozco Su{\'a}rez}, {Asensio
  Ramos}, \& {Trujillo Bueno}}]{2012ApJ...761L..25O}
{Orozco Su{\'a}rez}, D., {Asensio Ramos}, A., \& {Trujillo Bueno}, J. 2012,
  \apjl, 761, L25

\bibitem[{{Orrall} \& {Schmahl}(1976)}]{1976SoPh...50..365O}
{Orrall}, F.~Q. \& {Schmahl}, E.~J. 1976, \solphys, 50, 365

\bibitem[{Panesar {et~al.}(2013)Panesar, Innes, Tiwari, \& Low}]{panesar13}
Panesar, N.~K., Innes, D.~E., Tiwari, S.~K., \& Low, B.~C. 2013, A+A, 549, A105

\bibitem[{Parenti {et~al.}(2012)Parenti, Schmieder, Heinzel, \&
  Golub}]{Parenti12}
Parenti, S., Schmieder, B., Heinzel, P., \& Golub, L. 2012, ApJ, 754, 66

\bibitem[{Parenti \& Vial(2007)}]{Parenti07}
Parenti, S. \& Vial, J.-C. 2007, A+A, 469, 1109

\bibitem[{Schmahl \& Orrall(1986)}]{Schmahl86}
Schmahl, E. \& Orrall, F. 1986, NASA Conf. Publ., 2442, 127

\bibitem[{{Schmieder} {et~al.}(2014){Schmieder}, {Tian}, {Kucera}, {L{\'o}pez
  Ariste}, {Mein}, {Mein}, {Dalmasse}, \& {Golub}}]{2014A&A...569A..85S}
{Schmieder}, B., {Tian}, H., {Kucera}, T., {et~al.} 2014, \aap, 569, A85

\bibitem[{Su {et~al.}(2014)Su, G\"{o}m\"{o}ry, Veronig, Temmer, Wang,
  Vanninathan, Gan, \& Li}]{Su14}
Su, Y., G\"{o}m\"{o}ry, P., Veronig, A., {et~al.} 2014, ApJ Letters, 785, L2

\bibitem[{Su {et~al.}(2012)Su, Wang, Veronig, Temmer, \& Gan}]{Su12}
Su, Y., Wang, T., Veronig, A., Temmer, M., \& Gan, W. 2012, ApJ Letters, 756,
  L41

\bibitem[{{Warren} {et~al.}(2014){Warren}, {Ugarte-Urra}, \&
  {Landi}}]{Warren14}
{Warren}, H.~P., {Ugarte-Urra}, I., \& {Landi}, E. 2014, \apjs, 213, 11

\bibitem[{Wedemeyer {et~al.}(2013)Wedemeyer, Scullion, Rouppe van~der Voort,
  Bosnjak, \& Antolin}]{Wedemeyer13}
Wedemeyer, S., Scullion, E., Rouppe van~der Voort, L., Bosnjak, A., \& Antolin,
  P. 2013, ApJ, 774, 123

\bibitem[{Wiik {et~al.}(1993)Wiik, Dere, \& Schmieder}]{Wiik93}
Wiik, J., Dere, K., \& Schmieder, B. 1993, A+A, 273, 267

\bibitem[{{Young} {et~al.}(2012){Young}, {O'Dwyer}, \& {Mason}}]{Young12}
{Young}, P.~R., {O'Dwyer}, B., \& {Mason}, H.~E. 2012, \apj, 744, 14

\end{thebibliography}
\end{document}